\documentclass{article}
\usepackage[utf8]{inputenc}
\usepackage{amsmath}
\usepackage{physics}
\usepackage{subcaption}
\usepackage{graphicx}
\usepackage{amssymb}
\usepackage{braket}
\usepackage[affil-it]{authblk}
\usepackage{vmargin}
\usepackage{multicol}
\usepackage[colorlinks=true,citecolor=red,urlcolor=blue,linkcolor=blue]{hyperref}

\title{Vector Dark Matter from the $5-$Dimensional Representation of $SU(2)_L$}
\author[a,b]{Patricio Escalona\thanks{patricio.escalona@sansano.usm.cl}}
\author[a,b,c]{Sebasti\'an Acevedo\thanks{sebastian.acevedoe@sansano.usm.cl}}
\author[b]{Paulo Areyuna\thanks{paulo.areyuna@sansano.usm.cl}}
\author[a,c]{Gonzalo Benítez-Irarrázabal\thanks{gonzalo.benitezi@usm.cl}}
\author[a]{Pablo Solar\thanks{pablo.solar@usm.cl }}
\author[a,b,c]{Alfonso Zerwekh\thanks{alfonso.zerwekh@usm.cl}}
\affil[a]{Departamento de F\'isica, Universidad T\'ecnica Federico Santa Mar\'ia, Avenida España 1680, Valparaı\'iso, Chile}
\affil[b]{Millennium Institute for Subatomic Physics at High Energy Frontier – SAPHIR,
Fern\'andez Concha 700, Santiago, Chile}
\affil[c]{Centro Cient\'ifico-Tecnol\'ogico de Valpara\'iso, Universidad T\'ecnica Federico Santa Mar\'ia,
Avenida España 1680, Valpara\'iso, Chile}
\date{\today}

\begin{document}

\maketitle

\begin{abstract}
The introduction of electroweak multiplets that transform under any representation of the standard $SU(2)_L$ gauge group suggests the existence of electrically neutral stable particles capable of serving as cold dark matter in the $\Lambda$CDM cosmological model.
This paradigm, known as \textit{minimal dark matter}, has primarily focused on spin-$0$ and spin-$1/2$ particles.
We extend this study to the spin-1 case using the 5-dimensional real representation.
We address unitarity concerns arising from the model's interactions with electroweak and Higgs fields of the Standard Model, investigating implications for dark matter relic density, direct and indirect detection, including non-perturbative Sommerfeld enhancement for the latter.
Collider signatures of the proposed model are also examined.
Our findings suggest that the model remains consistent with experimental constraints, particularly for dark matter masses on the order of dozens of TeV, and could potentially be tested using $\gamma$-ray observatories such as CTA.
\end{abstract}

\section{Introduction}
The Standard Model of particle physics (SM) has been remarkably successful but remains incomplete.It fails to account for natural phenomena such as neutrino oscillations, the baryon asymmetry problem, and dark matter (DM).
The issue of DM is particularly pressing given the accumulated astrophysical and cosmological evidence supporting its existence and its crucial role in the formation of cosmic structure, as described by the $\Lambda$CDM model.
Among the various proposed candidates, Weakly Interacting Massive Particles (WIMPs) stand out as the simplest and most well-motivated due to their thermal production in the early universe \cite{Arcadi:2024ukq}.
The study of WIMPs has evolved significantly in recent decades, initially focusing on robust ultraviolet models such as the nearly degenerate supersymmetric higgsino DM \cite{Kowalska:2018toh}.
However, the absence of weak-scale non-standard phenomena has prompted a shift in research towards more phenomenologically grounded approaches, including effective theories \cite{Aebischer:2022wnl,Zaazoua:2021xls,Yamashita:2024krp}, simplified models \cite{DiazSaez:2021pfw,Banerjee:2021hfo,Yaguna:2021rds,Das:2022oyx,Belanger:2022esk,Cho:2023oad} and universal extra dimensions \cite{Maru:2018ocf}, among others.

The primary motivation behind these approaches is to provide a comprehensive explanation for DM phenomenology using minimal assumptions, aiming to minimize the number of parameters or degrees of freedom in the Lagrangian density while maintaining generality.
Models featuring colorless multiplets characterized by specific spins, hypercharges, and weak isospins are categorized as minimal DM models \cite{Cirelli:2005uq,Cirelli:2009uv,Cirelli:2015bda}. 
This framework offers a viable DM scenario with minimal parameters, typically including the tree-level mass of the multiplet and additional parameters for bosonic potentials.
Prominent candidates in these models are scalars or fermions with masses on the order of a few TeV, often represented in electroweak multiplets of dimensionality up to 5 or 7 \cite{Cirelli:2005uq}.
Vector candidates, however, encounter perturbative unitarity challenges when their mass parameter arises from a Proca term rather than from spontaneous local symmetry breaking.
Recent efforts have focused on spin-1 minimal DM in both fundamental and adjoint representations \cite{Saez:2018off,Belyaev:2018xpf}.

In this work, we extend the SM by incorporating a massive spin-1 field in the 5-dimensional representation of weak isospin.
The candidate for DM corresponds to the neutral component of the quintuplet, with its specific hypercharge chosen accordingly.
Radiative corrections contribute to the mass splitting among the quintuplet components, typically rendering the neutral component slightly lighter than the charged counterparts.
The Lorentz and gauge symmetries inherent in the model prevent tree-level decay modes in the Lagrangian density, ensuring DM stability due to an accidental symmetry.
As bosonic fields, these new particles interact renormalizably with the Higgs field and possess interactions beyond those induced by covariant derivatives in their kinetic terms.
We systematically analyze the violation of perturbative unitarity in longitudinally polarized scattering processes to delineate permissible parameter spaces for safe perturbative calculations.
Further constraints on the parameter space arise from ensuring that the neutral vector component satisfies DM relic density requirements.
The interplay of numerous coannihilation channels, non-minimal gauge interactions, and unitarity considerations for coupling constants collectively push the saturation mass to scales approaching dozens of TeV.
Upon identifying parameter space regions that satisfy relic density constraints, we investigate the model's detection prospects on multiple fronts: conducting a numerical scan for direct detection signatures, characterizing the non-perturbative Sommerfeld enhancement effect on photon flux from DM annihilation, and employing Monte Carlo simulations to explore $VVh$ production at the FCC-hh.

The paper is organized as follows: in Sec.~\ref{sec: The model} we define the model.
In Sec.~\ref{sec: Perturbative Unitarity} we study the parameter values that respect perturbative unitarity in longitudinal vector scattering.
In Sec.~\ref{sec: The mass splitting} the radiative correction of the mass difference between components of different electric charges are presented.
In Sec.~\ref{sec: Relic density} we present the calculations of DM abundance in the early universe.
In Sec.~\ref{sec: Phenomenology} we study the phenomenological consequences of the model, performing a numerical scan based on the differential evolution algorithm to confront relic density predictions with direct detection in Sec.~\ref{subsec: Direct}, studying the impact of the Sommerfeld enhancement on indirect searches solving the non-relativistic Schrodinger equation for DM annihilation in Sec.~\ref{subsec: Indirect}, and exploring the scope of the model in future colliders in Sec.~\ref{subsec: Collider}.
Finally in Sec.~\ref{sec: Conclusions} we state our conclusions.
In Appendix~\ref{sec: Appendix A} and Appendix~\ref{sec: Appendix Sommerfeld}, we present supplementary information about scattering amplitudes for longitudinally polarized processes and the calculation of the Sommerfeld enhancement for this model.

\section{The Model}
\label{sec: The model}
We supplement the SM Lagrangian density by introducing a weak isospin multiplet $V_\mu^i$ transforming under the 5-dimensional real representation of $SU(2)_L$.
To achieve this, we construct a field strength tensor for the new vector bosons, implementing minimal substitution of spacetime derivatives with covariant ones, and include all the terms consistent with the symmetries.
For an arbitrary hypercharge $Y$, the explicit form of these derivatives is
\begin{equation}
    D_\mu V_\nu^i=\left(\partial_\mu \delta^{ij} + \frac{ig}{2} (T^a)^{ij}W_\mu^a+\frac{ig' Y}{2}\delta^{ij}B_\mu\right)V_\nu^j.
\end{equation}
Then, we construct the field strength tensor as
\begin{equation}
    V_{\mu \nu}^i = D_\mu V_\nu^i - D_\nu V_\mu^i,
\end{equation}
where $i,j=1,\dots,5$ and $T^a$ are the generators of the 5-dimensional representation of $SU(2)_L$ \cite{Cai:2017fmr},
\begin{equation}
    T^{1}=\frac{1}{2}
    \begin{pmatrix}
        \phantom{0} & -2 & \phantom{0} & \phantom{0} & \phantom{0} \\
        -2 & \phantom{0} & -\sqrt{6} & \phantom{0} & \phantom{0} \\
        \phantom{0} & -\sqrt{6} & \phantom{0} & \sqrt{6} & \phantom{0} \\
        \phantom{0} & \phantom{0} & \sqrt{6} & \phantom{0} & 2 \\
        \phantom{0} & \phantom{0} & \phantom{0} & 2 & \phantom{0}
    \end{pmatrix},
    \quad
    T^{2}=\frac{1}{2i}
    \begin{pmatrix}
        \phantom{0} & -2 & \phantom{0} & \phantom{0} & \phantom{0} \\
        2 & \phantom{0} & -\sqrt{6} & \phantom{0} & \phantom{0} \\
        \phantom{0} & \sqrt{6} & \phantom{0} &\sqrt{6} & \phantom{0} \\
        \phantom{0} & \phantom{0} & -\sqrt{6} & \phantom{0} & 2 \\
        \phantom{0} & \phantom{0} & \phantom{0} & -2 & \phantom{0}
    \end{pmatrix},
\end{equation}
\begin{equation}
    T^{3}=\mathbf{diag} (2,1,0,-1,-2).
\end{equation}
Note that the exotic quintuplet is not a gauge field; hence, it transforms homogeneously under weak global transformations: $V'_\mu \rightarrow U V_\mu $ with $U\in SU(2)_L$.
Therefore, its field strength tensor lacks an inhomogeneous term proportional to the group structure constants.
The proposed Lagrangian density is
\begin{equation}
    \mathcal{L}=\mathcal{L}_{SM}+\mathcal{L}_{Proca} + \mathcal{L}_{NM},
\end{equation}
where additionally to the SM we add $\mathcal{L}_{Proca}$ corresponding to
\begin{equation}
    \mathcal{L}_{Proca}=-\frac{1}{4} 
    \qty(V_{\mu \nu} ^i)^\dagger V^{\mu \nu i} + \frac{1}{2} m^2_V \qty(V_\mu^i)^\dagger V^{\mu i}.
\end{equation}
Furthermore, we incorporate non-minimal terms involving interactions between the new fields and the $W$ and $B$ field strength tensors, a `gauge-fixing-like' term \cite{VanDong:2021xws}, a Higgs portal interaction, and the most general four-point self-interaction for $V$, respectively, as displayed in the following expression:
\begin{equation}
    \begin{split}
        \mathcal{L}_{NM}=\,&i\kappa gV^{\dagger\mu}W_{\mu\nu}V^{\nu}+ iY\kappa' g'V^{\dagger\mu}B_{\mu\nu}V^{\nu}+\frac{1}{\xi}(D_{\mu}V^\mu)^\dagger(D_\nu V^{\nu}) + \frac{\lambda_{HV}}{2}(\Phi^\dagger \Phi)(V_{\mu}^\dagger V^{\mu})\\
        &+\alpha_1\left(V_\mu^\dagger V^\mu \right)\left(V_\nu^\dagger V^\nu \right)+\alpha_2\left(V_\mu^\dagger V^\nu \right)\left(V_\nu^\dagger V^\mu \right)+\alpha_3\left(V_\mu^\dagger V^\nu \right)\left(V^{\dagger\mu} V_\nu \right).
    \end{split}
    \label{lagrangian_nm}
\end{equation}
The constants $\kappa$ and $\kappa'$ modulate the non-minimal interactions with the SM gauge bosons.
These parameters along with $1/\xi$ are considered free parameters, but they play a crucial role in a UV completion of the model with an extended gauge sector and  have important consequences for perturbative unitarity, as detailed in in Sec.~\ref{sec: Perturbative Unitarity}.

Moreover, interaction with the Higgs doublet alters the tree-level mass of the new particles after spontaneous symmetry breaking, resulting in the physical mass $M_V^2=m^2_V+\frac{1}{2}\lambda_{HV}v^2$.
Additionally, the term proportional to $1/\xi$ modifies the propagator of the new vector fields as:
\begin{equation}
    \Pi_{\mu\nu}(q)=\dfrac{-i}{q^2-M_V^2}\left(g_{\mu\nu}-\dfrac{(1-\xi)q_\mu q_\nu}{q^2-\xi M_V^2} \right).
\end{equation}

The assignment of hypercharge of the new degrees of freedom determines which component is electrically neutral, and thus is a DM candidate.
Setting $Y=0$ is particularly appealing because it suppresses tree-level DM-nucleon electroweak interactions mediated by $Z$ bosons, keeping the model viable under the strict upper bounds on the WIMP-nucleon direct detection cross-section. 
The direct signals from this model are Higgs mediated at tree-level. 
Thus, our setup identifies the DM candidate as the third component of the quintuplet, with weak isospin quantum number $0$ and zero hypercharge.
Importantly, this choice for the hypercharge completely cancels the term proportional to $\kappa'$.

The implementation of the model was performed using \verb|FeynRules| \cite{Christensen:2008py,Alloul:2013bka}.

\section{Perturbative Unitarity}
\label{sec: Perturbative Unitarity}
Incorporating additional massive vector fields typically results in violations of perturbative unitarity in scattering processes involving these spin-1 bosons.
This issue manifests itself as scattering amplitudes that do not decrease sufficiently fast with increasing $s=E_{CM}^2$.
Three problematic categories of terms appear in the large $s$ expansion of scattering amplitudes: those proportional to $s^2$, terms linear in $s$, and those independent of $s$.

In the SM, the cancellation of quadratic terms is automatic due to the gauge structure, while the vanishing of linear terms occurs when the Higgs-gauge boson coupling has the form mandated by gauge invariance. 
Moreover, the $s$-independent term imposes an upper bound on the Higgs mass \cite{Lee:1977eg}.
In contrast, for spin-1 bosons transforming in the adjoint representation of weak isospin \cite{Belyaev:2018xpf}, the quartic terms cancel out, but the linear terms do not, thereby imposing an upper bound for the energy scale of validity of the model.
In our case, the quadratic terms do not vanish because the new massive vectors are not gauge bosons; therefore, their self-interactions do not arise from a Yang-Mills term. 

To address this issue, we compute the scattering amplitudes considering general values for the coupling constants and study the parameter space that ensures perturbative unitarity (see, e.g., \cite{Kahlhoefer:2015bea,Abe:2020mph,Gopalakrishna:2016tku,ElHedri:2014zng,Cynolter:2015sua,Lebedev:2011iq,Barman:2021qds,Englert:2016joy,Milagre:2024wcg}).
We examined the partial-wave expansion of the scattering amplitude for all possible processes including the longitudinal components of both the new $V$ bosons and the massive SM bosons.
The most stringent unitarity condition arises from the first term ($l=0$) of the partial wave expansion:
\begin{equation}
    a_0(s)=\dfrac{1}{32\pi}\displaystyle \int_{-1}^1 i\mathcal{M}\, d(\cos{(\theta)}).
\end{equation}

Let us consider the $V_LV_L\to V_LV_L$ process, where $V$ represents any component of the multipĺet: $V^{\pm\pm}, V^\pm$ or $V^0$, and the subscript $L$ denotes longitudinal degrees of freedom.
Explicit expressions can be found in the Appendix~\ref{sec: Appendix A}. 
All problematic quadratic terms share the structure $\sim (A \alpha_1 + B\alpha_2 + A  \alpha_3 + C \kappa^2 g^2) s^2$, where $A, B, C$ are different integers specific to each process.
Note that the $\alpha_i$ couplings modulate the four-point self-interactions of $V$ particles, while $\kappa$ arises from non-minimal interactions with SM gauge bosons.
Given that $V$ self-interactions are not affected by isospin quantum numbers, unlike interactions mediated by SM vector bosons, the only way to cancel all of these terms simultaneously is to set $\alpha_1= -\alpha_3, \alpha_2=0$ and $\kappa=0$.  Under these conditions and applying the constrain derived from the optical theorem, $|a_0(s)|\leq 1$, the more restrictive scale of unitarity violation for $V_LV_L\to V_LV_L$ scattering, derived from the linear term in $s$, is
\begin{equation}
    \Lambda \approx \dfrac{8\sqrt{\pi} M_V^2}{|\lambda_{HV}| v}.
\end{equation}
This energy scale exhibits a similar structure to that obtained for a vector triplet \cite{Belyaev:2018xpf}. 
These conditions impose severe constraints on the structure of $V$ self-interactions.

The partial wave amplitude for the $ZZ \to V^0V^0$ process features only a linear term in $s$. This process imposes a more restrictive unitarity violation scale, which is independent of $\kappa$ and the quartic couplings. Following a standard procedure, we determine that the unitarity violation scale for this process is
\begin{equation}
    \Lambda'\approx\dfrac{4 \sqrt{2\pi} M_V}{\sqrt{|\lambda_{HV}|}}.
\end{equation}
\begin{figure}[t]
    \centering
    \includegraphics[width=0.49\textwidth]{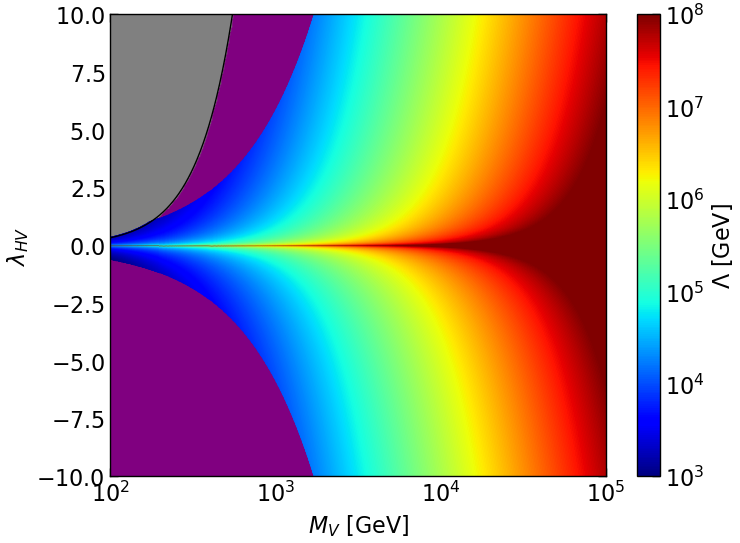}
    \includegraphics[width=0.49\textwidth]{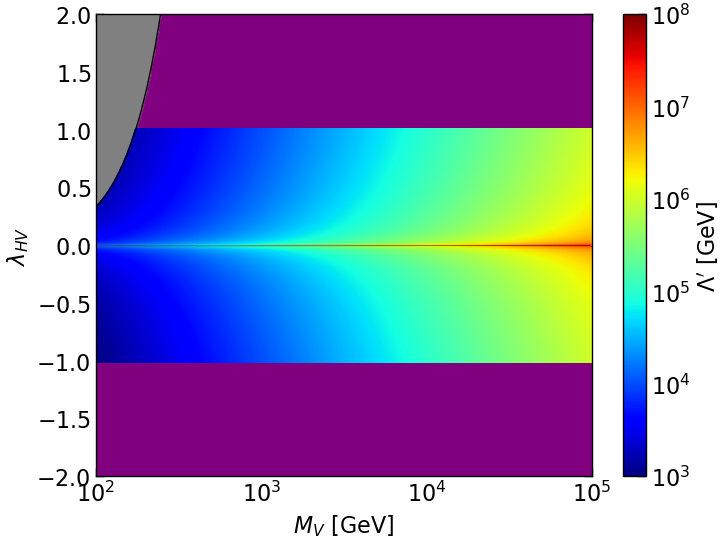}
    \caption{Maximum energy scale allowed by unitarity bounds as a function of $\lambda_{HV}$ and $M_V$. The purple regions represent the zones where $\Lambda, \Lambda' < 10 M_V$. The gray region is forbidden as the Proca mass coefficient $m_V$ becomes imaginary. Left: Unitarity violation scale $\Lambda$ on the $VV\to VV$ amplitude. Right: Unitarity violation scale $\Lambda'$ on the $ZZ\to V^0V^0$ amplitude.}
    \label{scale_VVVV}
\end{figure}
Next, we compare the restrictions on the parameter space imposed by the values of $\Lambda$ and $\Lambda'$.
The left panel of Fig.~\ref{scale_VVVV} displays $\Lambda$ as a color map, as a function of $\lambda_{HV}$ and $M_V$.
Similarly, the right panel shows $\Lambda'$.
The horizontal axis in these plots represents the tree-level mass of the $V$ bosons, which, as discussed in Sec.~\ref{sec: Relic density}, we consider in the regime of a few tens of TeV.
Additionally, we require that the unitarity violation scale be at least of the order of $10 M_V$ to ensure reliable non-relativistic perturbative calculations for DM phenomenology  and collider phenomenology at the FCC.
From Fig.~\ref{scale_VVVV} it becomes evident that the most stringent constraint on $\lambda_{HV}$ in the mass regime relevant for DM phenomenology arises from $\Lambda'$, highlighted in the purple regions.

Up to this point, our analysis has been independent of the coupling $1/\xi$.
Now we shift our focus to $2\to2$ scattering processes involving the non-minimal interaction between the new vector bosons and the SM gauge bosons originating from the `gauge-fixing-like' term.
Unfortunately, in this scenario, it is not feasible to cancel out all the terms proportional to $s^2$.
Nevertheless, we have confirmed that the condition $|a_0|<1$ remains satisfied if, in addition to the aforementioned conditions, we select $\xi$ within the interval $[-2/3, -0.1]$.

Finally, considering unitarity constrains the Lagrangian density is constrained to take the form
\begin{equation}
    \begin{split}
    \mathcal{L}=\,&\mathcal{L}_{SM}-\frac{1}{4} 
    \qty(V_{\mu \nu} ^i)^\dagger V^{\mu \nu i} + \frac{1}{2} m^2_V \qty(V_\mu^i)^\dagger V^{\mu   i}+\frac{1}{\xi}(D_{\mu}V^\mu)^\dagger(D_\nu V^{\nu})\\&+\frac{\lambda_{HV}}{2}(\Phi^\dagger \Phi)(V_{\mu}^\dagger V^{\mu})  +\alpha\left[\left(V_\mu^\dagger V^\mu \right)\left(V_\nu^\dagger V^\nu \right)-\left(V_\mu^\dagger V^\nu \right)\left(V^{\dagger\mu} V_\nu \right)\right],
    \end{split}
    \label{final_lagrangian}
\end{equation}
where $|\alpha_1|=|\alpha| \lesssim 4\pi$. $|\lambda_{HV}|\lesssim 4\pi$ and $1/\xi \in [-10,-3/2]$. 

Note that the amplitudes of longitudinally polarized scattering presented in Appendix~\ref{sec: Appendix A} are calculated at tree level. In general, these amplitudes receive radiative corrections, so the choices of constants previously stated are corrected order by order in perturbation theory. It has been verified at the one-loop level that these corrections are at most logarithmic, thus supporting consistency as a low-energy effective theory.

%When the Lagrangian parameters are corrected at one loop, the coupling constants involved in unitarity ($\alpha_i$, $\xi$, $\kappa$ and $\lambda_{HV}$) receives a correction from diagrams that have propagators of the form $\frac{1}{k^2-M^2}$, that may be scalars or vectorial. The superficial degree of divergence are $\text{ln} \Lambda$, $\Lambda^{-2}$ and $\Lambda^{-4}$ for diagrams with 2, 3 and 4 bosonic propagators respectively. Therefore, running this parameters below the scale $\Lambda$ and $\Lambda'$, since the correction will be logarithmic, does not generate $s^2$ contribution that violates unitarity.

\section{Mass Splitting}
\label{sec: The mass splitting}
The tree-level mass of the new vector bosons, after the Higgs field develops a vacuum expectation value, is given by
\begin{equation}
    M_V^2=m^2_V+\dfrac{\lambda_{HV} v^2}{2},
    \label{mass_vector} 
\end{equation}
where $M_V$ is the physical mass, $m_V$ is the coefficient of the Proca bilinear term, $\lambda_{HV}$ is the Higgs portal coupling, and $v$ is the vacuum expectation value of the Higgs.
In the lowest order of perturbation theory, there is a total degeneracy among the masses of the components of the vector quintuplet\footnote{In principle, we could consider the following additional interactions: 
\begin{equation}
    \mathcal{L} \supset \lambda_{H V}^{\prime}\left(V^{\dagger} T_V^a V\right)\left(H^{\dagger} T_H^a H\right),
\end{equation}
which introduces a mass difference in the components at tree level.
For simplicity, we assume that the coupling $\lambda_{HV}'$ is sufficiently small to have negligible effects on the calculation.}.
However, the radiative corrections for the charged and neutral components differ, primarily due to the electromagnetic and non-minimal interactions present in the charged components but absent in the neutral components. 
Consequently, the neutral component emerges as the lightest among the quintuplet.
Moreover, it's noteworthy note that the appearance of $V_\mu$ fields in pairs in the Lagrangian implies an accidental $Z_2$ symmetry.
These conditions collectively ensure the stability of the neutral component, endowing it with DM properties.

\begin{figure}[t]
    \centering
    \includegraphics[width=0.8\textwidth]{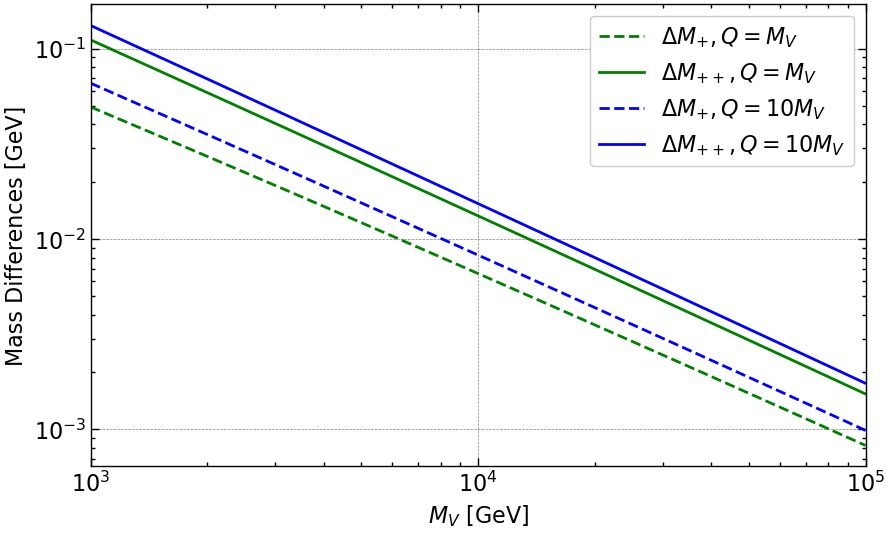}
    \caption{Mass splitting between the components of the spin-1 weak isospin quintuplet as a function of the tree-level mass $M_V$. The solid lines represent the difference between the doubly-charged and single-charged components, $\Delta_{++}=M_{V^{\pm\pm}}-M_{V^{\pm}}$, while the dashed lines represent the difference between the single-charged and neutral components, $\Delta_{+}=M_{V^{\pm}}-M_{V^{0}}$. The plot includes two cases: $Q=M_V$ and $Q= 10 M_V$. The parameters used are $\kappa=0, \alpha_1=-\alpha_3=0.1, \xi^{-1}=-1.5$ and $\lambda_{HV}=0.1$.}
    \label{splitting}
\end{figure}

We adopt the methods outlined in \cite{Belyaev:2018xpf} to compute the mass differences. 
This involves evaluating the real, finite, and transverse parts of the one-loop self-energies ($\Sigma^{++,+,0}$) for both charged and neutral components.
Divergent integrals are  handled using dimensional regularization\footnote{This procedure introduces the renormalization scale $Q$, which is fixed in Fig.~\ref{splitting}.} within the modified minimal subtraction scheme, under the approximation $M_V \gg m_{W,Z}$. 
The self-energies and resulting mass differences are related according to
\begin{equation}
\Delta M_{+}=M_V \sum_{n=1}^{\infty}(-1)^n\left(\begin{array}{c}
\frac{1}{2} \\
n
\end{array}\right)\left[\left(\frac{\Sigma^{+}\left(M_V^2\right)}{M_V^2}\right)^n-\left(\frac{\Sigma^0\left(M_V^2\right)}{M_V^2}\right)^n\right].
\end{equation}
where the binomial coefficient is
\begin{equation*}
    \binom{\frac{1}{2}}{n}=\frac{1}{n!}\left(\frac{1}{2}\right)\left(\frac{1}{2}-1\right)\left(\frac{1}{2}-2\right) \ldots\left(\frac{1}{2}-n+1\right).
\end{equation*}
Similarly, we express $\Delta M_{++}$ in terms of differences of powers of $\Sigma^{++}$ and $\Sigma^{+}$.
The mass difference is truncated in the term $n = 1$.

The results are shown in Fig.~\ref{splitting}. 
Clearly, the mass splitting remains relatively small in the regime of large $M_V$.
Despite initial expectations of a mass difference around $100$ MeV, as observed in the adjoint representation, the presence of non-minimal terms leads to a monotonically decreasing behavior in the quintuplet case.

The calculation of the self-energies was carried out using the \verb|FeynArts| \cite{Hahn:2000kx} and \verb|FeynCalc| \cite{Mertig:1990an} packages within the \verb|Mathematica| environment.

\section{Relic Density}
\label{sec: Relic density}
To gain insight into the impact of various parameter configurations on DM relic abundance, we conducted an initial numerical analysis.
A detailed parameter scan is outlined in Sec.~\ref{subsec: Direct}, which also covers direct detection signals.
Our calculations used \verb|micrOMEGAs 5.3.41| \cite{Belanger:2010pz,Alguero:2023zol} within the freeze-out scenario.
It is crucial to note that DM production hinges on the mass splitting among the vector components, necessitating consideration of how mass differences vary with model parameters when solving the Boltzmann equation.
However, as demonstrated in the previous section, the $\Delta M_{+,++}$ remains relatively small, typically within the range of hundreds of MeV, which is significantly less than the mass of the DM candidate assumed at the WIMP scale.
\begin{figure}[t]
    \centering
    \includegraphics[width=0.7\textwidth]{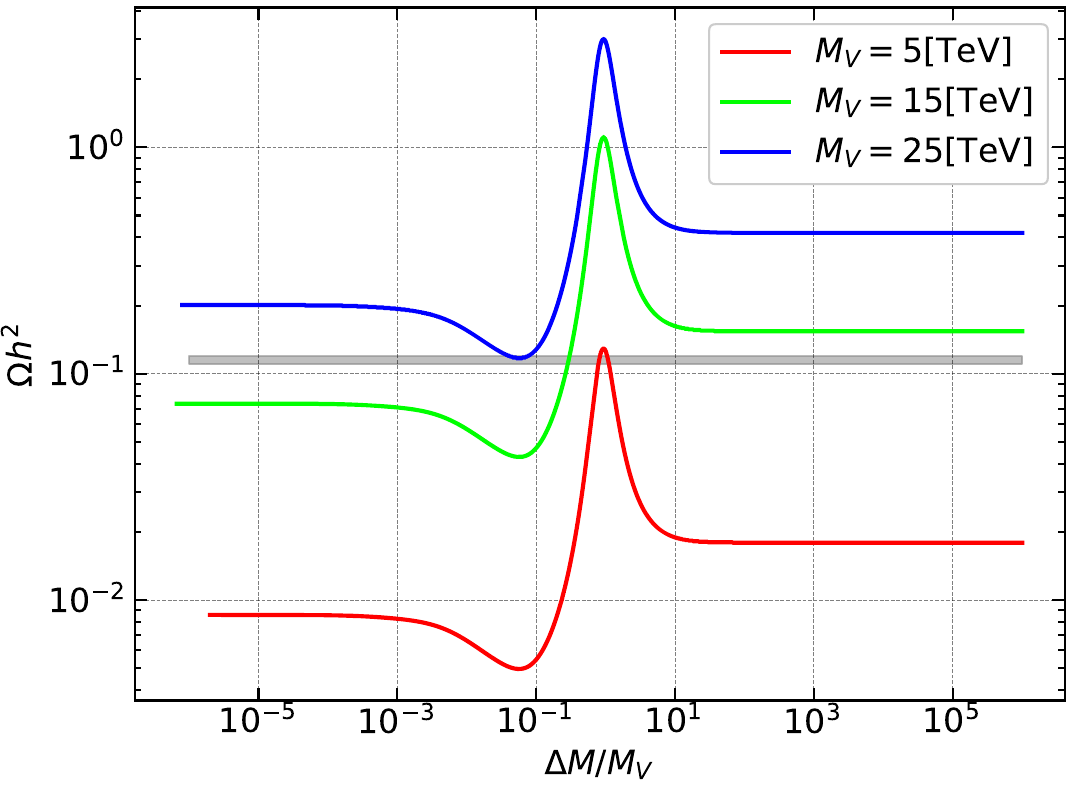}
    \caption{Relic density as a function of the ratio between the mass splitting $\Delta M_+$ and DM mass $M_V$. This plot  considered values for $1/\xi=-1.5$ and $\lambda_{HV}=-0.1$. The gray band represents the measured value of the relic density $\Omega h^2\sim 0.12$ \cite{Planck:2018vyg}. Different colored lines correspond to various DM mass values: $M_V = 5$ TeV (red), $M_V = 15$ TeV (green), and $M_V = 25$ TeV (blue).}
    \label{relic_vs_deltam}
\end{figure}
In Fig.~\ref{relic_vs_deltam}, we present the relic density as a function of the mass differences. 
Notably, the plot exhibits asymptotically constant behavior for $\Delta M/M_V \ll 1$, emphasizing the maximal impact of coannihilations in the quasi-degenerate limit.
Given our specific parameter values, we can approximate the mass splitting's exact value to be negligible for practical purposes.

We then examined how the model parameters influence the relic density and identifies the range that predicts the correct amount of DM in the present universe.
\begin{figure}[t]
    \centering
    \includegraphics[width=0.48\textwidth]{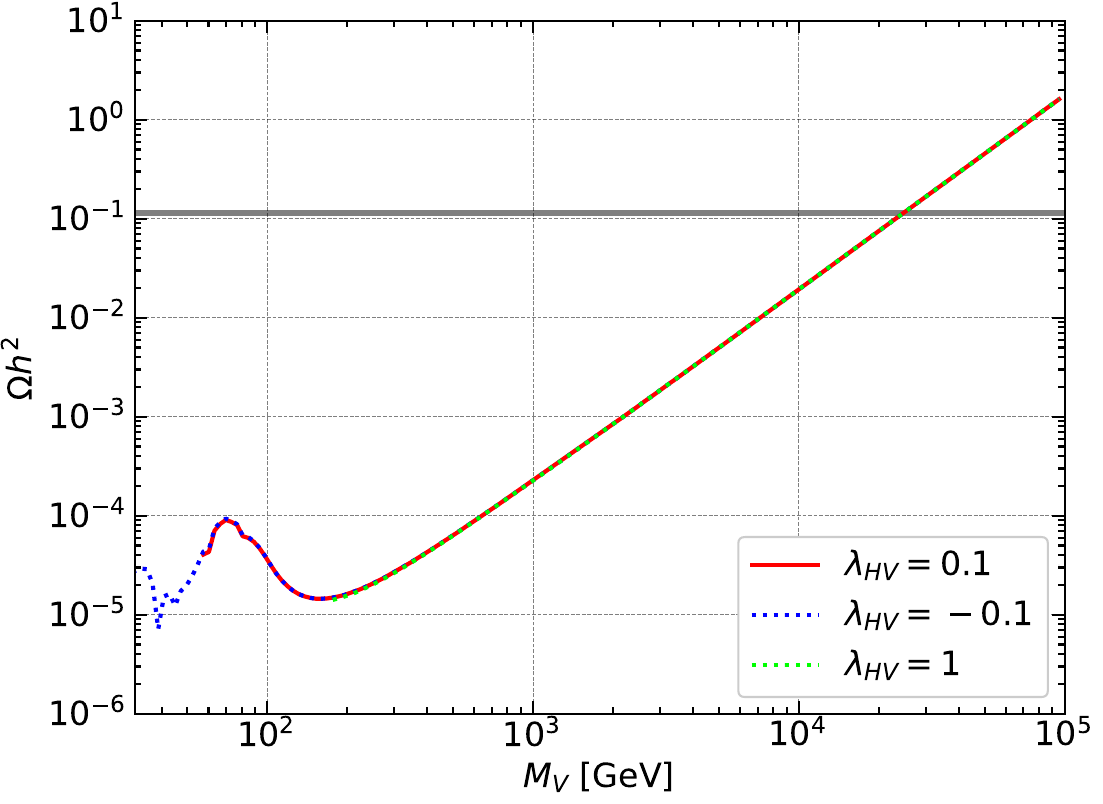}
    \includegraphics[width=0.48\textwidth]{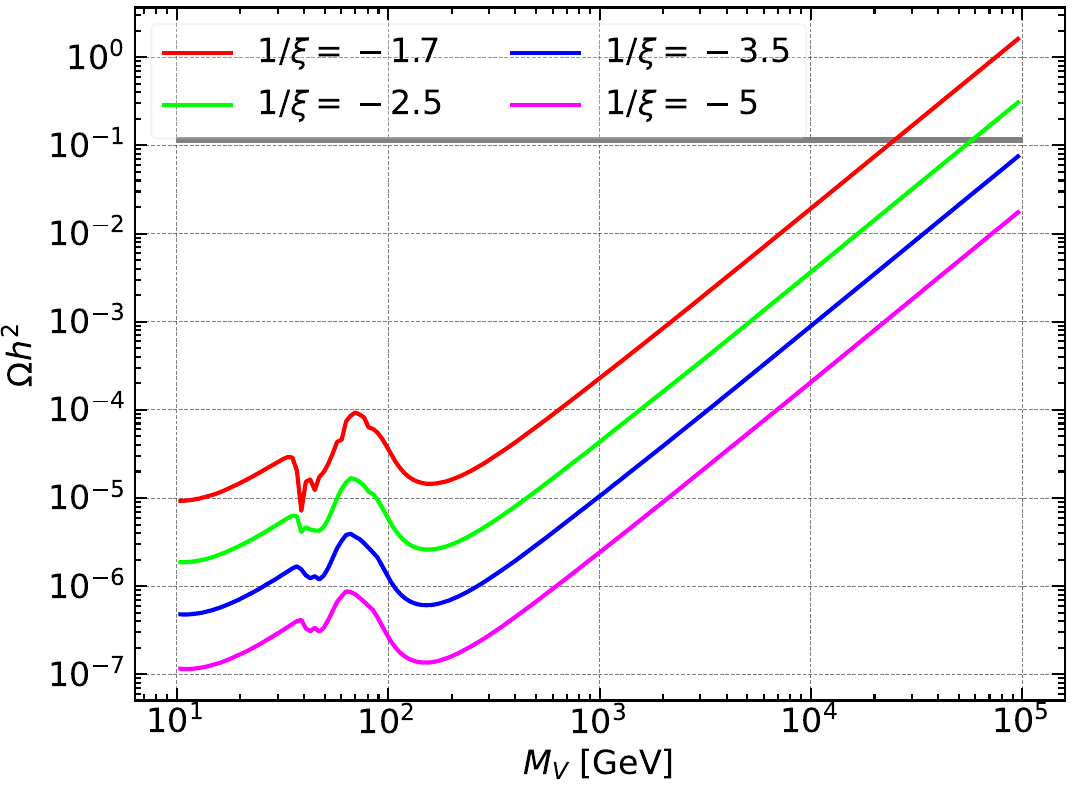}
    \caption{Relic density as a function of the vector mass for different choices of $\lambda_{HV}$ and $1/\xi$. These plots were obtained considering a fixed value of $\Delta M=10$ MeV. The solid gray bands correspond to the $3\sigma$ confidence interval given by the Planck mission \cite{Planck:2018vyg}. In the left plot, we take $\xi^{-1}=-1.7$. On the right plot, we take $\lambda_{HV}=-0.1$.}
    \label{relic_vs_mass}
\end{figure}
As shown in Fig.~\ref{relic_vs_mass}, there exists a minimum value of $M_V$ for each $\lambda_{HV}>0$ to prevents the Proca coefficient $m_V^2$ from becoming negative and ensuring the theory remains well-defined. 
Conversely, this lower limit is absent when $\lambda_{HV}<0$, allowing for a lower vector mass. 
In this cases, the electroweak resonances enhance the annihilation cross-section, creating noticeable dips in the relic density plot, as expected for WIMP scenarios of this type \cite{Saez:2018off,Belyaev:2018xpf}.
It's important to note that Higgs-mediated interactions have minimal impact on relic density calculations due to their weaker contributions and lesser influence on scattering amplitudes compared to charged-mediated processes involved in coannihilations.

An interesting observation is the substantial suppression of the relic density, achieving the measured value of $\Omega h^2=0.12$ at tens of TeV.
This strong suppression is related to the combined influence of numerous coannihilation channels, which significantly contribute to the depletion of the dark sector during the decoupling process, along with the non-minimal interactions between gauge bosons and the vector quintuplet.
It's noteworthy that the saturation mass can potentially be adjusted with a smaller value of $|1/\xi|$.
However, as previously mentioned, this parameter is constrained perturbative unitarity considerations, preventing saturation below $M_{V}\approx 20$ TeV.

\section{Phenomenology}
\label{sec: Phenomenology}
\subsection{Direct Detection and Numerical Scan}
\label{subsec: Direct}
To systematically explore the phenomenological aspects of the model in the regimes studied in previous sections, we conducted a random scan of the parameter space using the optimization method known as differential evolution \cite{Storn:1997uea}, implemented in the \verb|scipy| package for \verb|Python| \cite{Virtanen:2019joe}.
During this procedure, we calculated the relic density for each point using \verb|micrOMEGAs|, aiming to match the relic abundance inferred from the Planck satellite's CMB data. Specifically, we minimized a logarithmic likelihood function defined in accordance the measurement $\Omega h^2 = 0.12$ and its error bars.

The values of the coupling constants used in this scan were chosen to align with the unitary regime described in Sec.~\ref{sec: Perturbative Unitarity}, while the mass range was motivated by the relic density results discussed in Sec.~\ref{sec: Relic density}. The specific values are presented in Table~\ref{table1} along with the algorithm parameters.
\begin{figure}[t]
    \centering
    \includegraphics[width=0.7\textwidth]{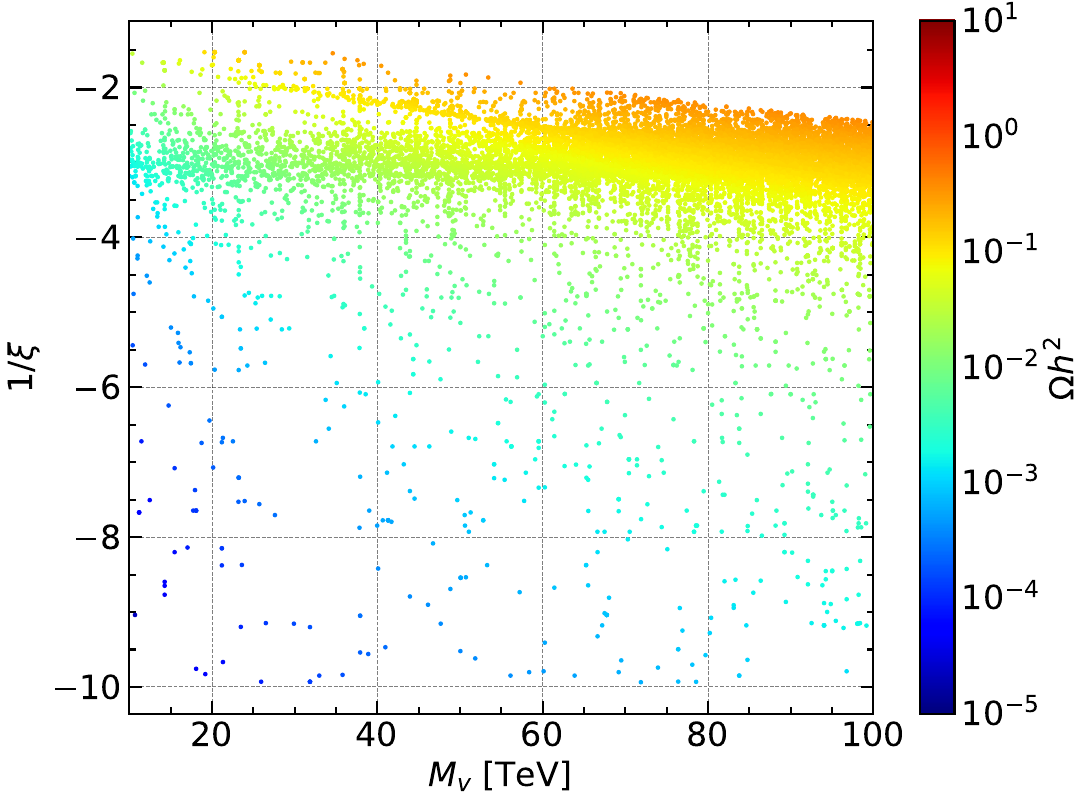}
    \caption{Random scan of the relic density as a function of the DM mass (horizontal axis) and the non-minimal $1/\xi$ parameter (vertical axis). The color map indicates the calculated relic density values obtained by solving the Boltzmann equation.}
    \label{diff_evol_relic}
\end{figure}
The scan results are shown in Fig.~\ref{diff_evol_relic}.
In this figure, the relic density is represented by colors as a function of the DM mass and the parameter $1/\xi$.
A correlation between the value of $\xi$ and the DM abundance is observed, as anticipated in Sec.~\ref{sec: Relic density}.
It should be noted that the unitarity region of $\xi$ can accommodate the measured abundance of DM.
Naturally, the relic density increases with the DM mass.
Additionally, the scan confirmed that the Higgs portal coupling $\lambda_{HV}$ has a subdominant influence compared to $\xi$ in the Boltzmann equations for DM production in the early universe. 

A common feature of WIMPs is that their effective interactions with nucleons can lead to potentially observable scattering processes, a strategy known as direct detection.
In our model, the Higgs portal induces a non-zero elastic scattering at tree level with quarks mediated by a Higgs boson in the t-channel.
A different choice of hypercharge for the multiplet would result in the exchange of a $Z$ boson in this channel; however, this scenario was not explored in this work.
\begin{figure}[t]
    \centering
    \includegraphics[width=0.49\textwidth]{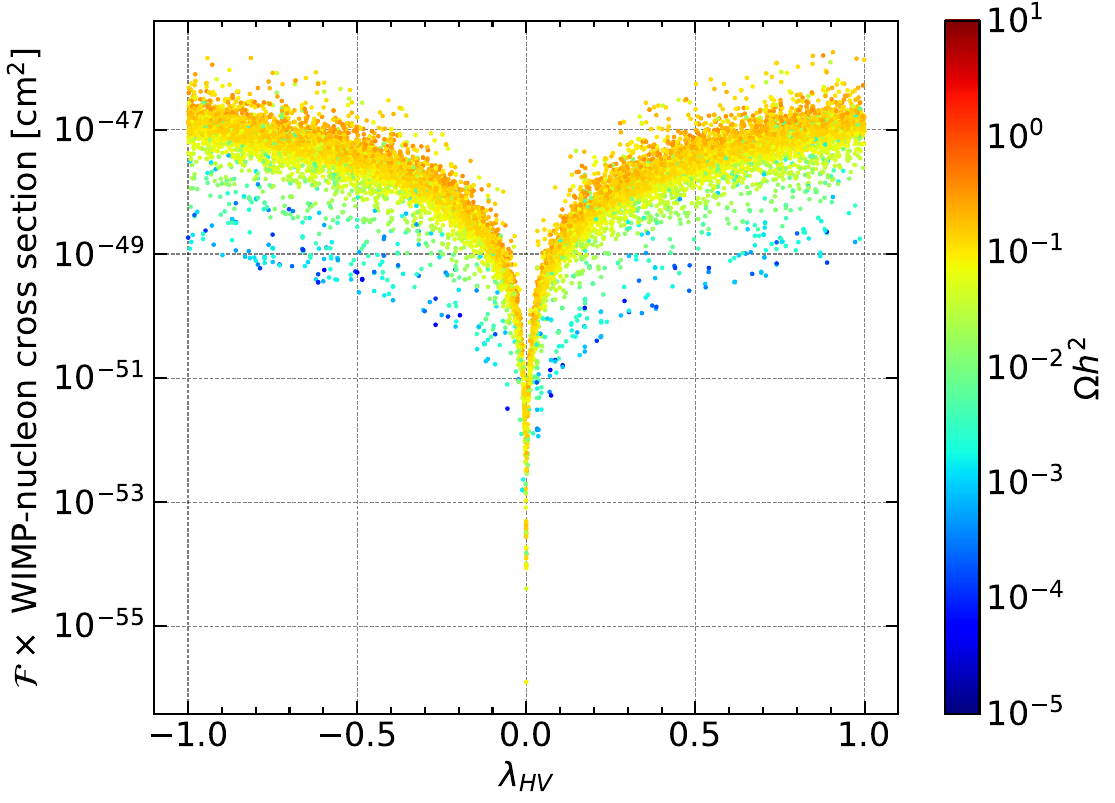}
    \includegraphics[width=0.49\textwidth]{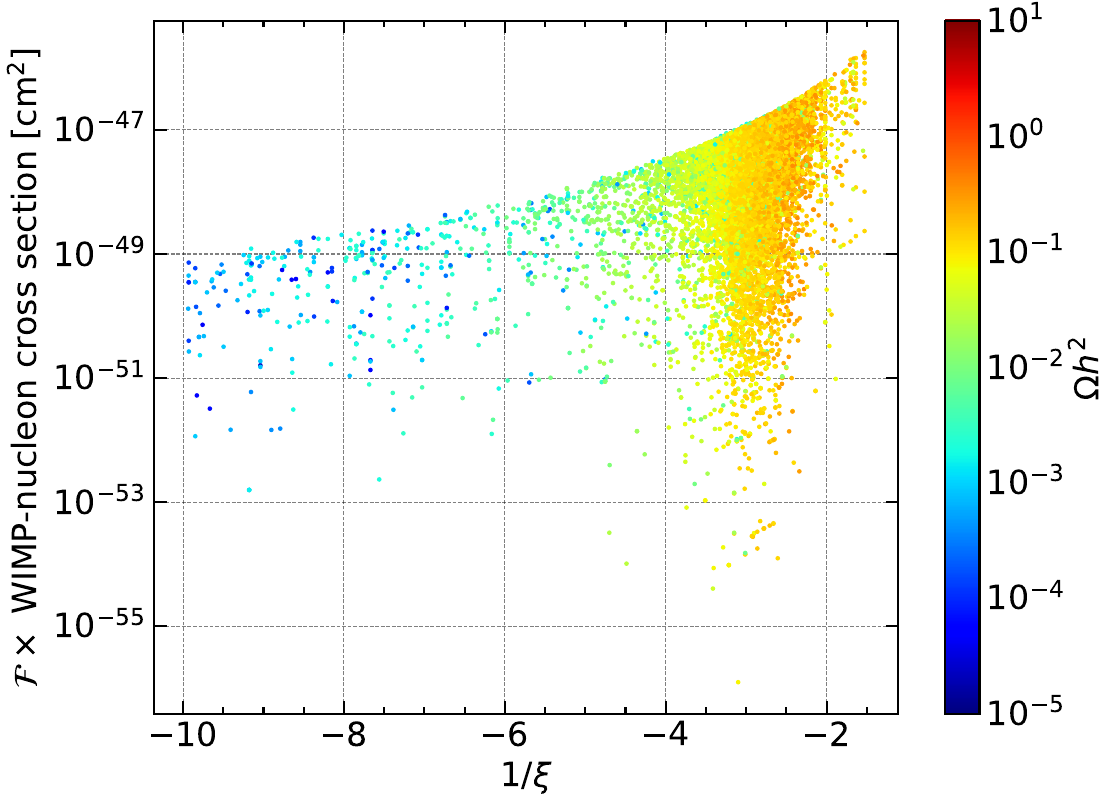}
    \caption{Spin-independent elastic DM-nucleon cross-section of the scanned points as a function of the coupling constants of the model. The color map represents the corresponding relic abundance. On the left plot, we show the Higgs portal dependence and on the right one we show the non-minimal $1/\xi$ dependence.}
    \label{DD_couplings}
\end{figure}

In Fig.~\ref{DD_couplings}, we project the spin-independent cross-section between DM and nucleons, weighted by the factor $\mathcal{F}$ defined as
\begin{equation}
    \mathcal{F}(1/\xi, \lambda_{HV},M_V)=\frac{\Omega h^2(1/\xi, \lambda_{HV},M_V)}{\Omega h^2_{PLANCK}},
\end{equation}
which quantifies the relative contribution of the model to the DM abundance measured by Planck.

In Fig.~\ref{DD_couplings}, the left panel shows this quantity as a function of $\lambda_{HV}$, while the right panel shows it as a function of $1/\xi$.
The symmetry of the graph on the left reflects the dependence of the cross-section on $\lambda_{HV}^2$.
The correlation between $1/\xi$ and direct detection observed in the graph on the right mirrors that shown in Fig.~\ref{diff_evol_relic}, as it enters through the factor $\mathcal{F}$.
Considering these observations, we note that direct detection prospects are below the current upper bounds on the DM-nucleon cross-section \cite{XENON:2018voc}, and even below the expected sensitivity for XENONnT \cite{XENON:2020kmp}, an experiment expected to impose one of the most stringent constraints on this signal.
\begin{figure}[t]
    \centering
    \includegraphics[width=0.7\textwidth]{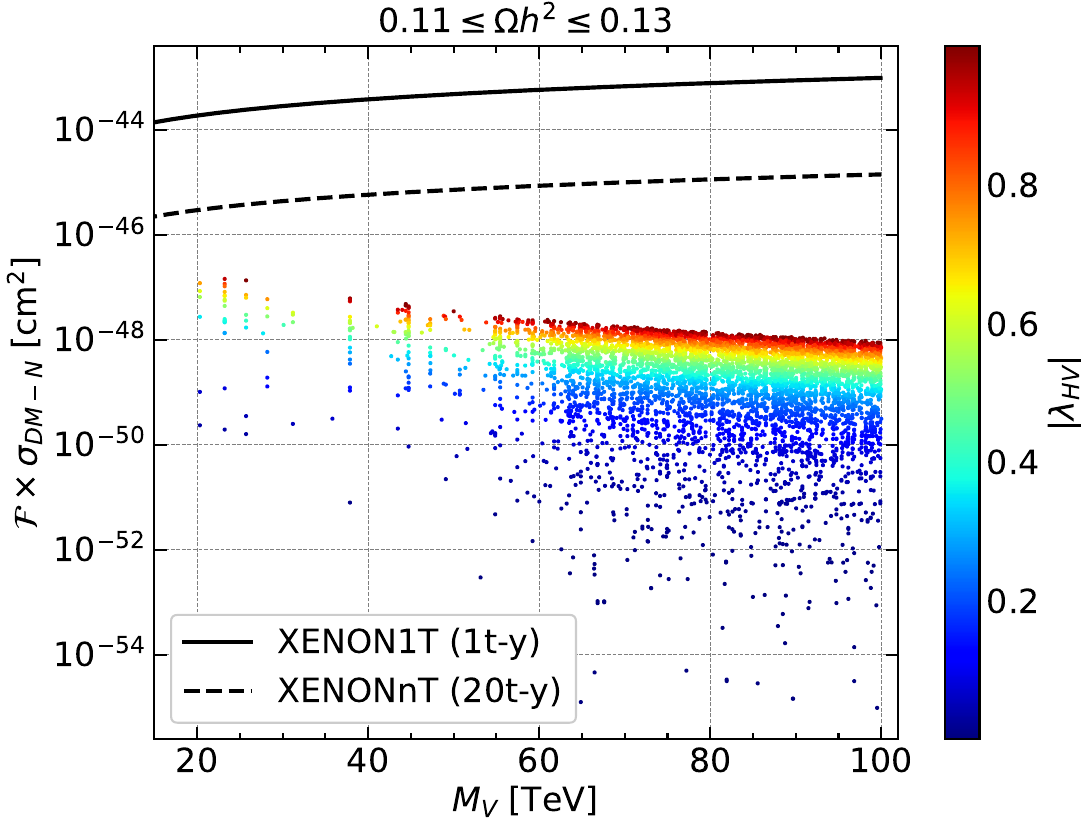}
    \caption{Effective spin-independent cross-section of the interaction between DM and nucleons as a function of the multiplet tree-level mass.  The solid black line represents the current upper exclusion limits from XENON1T, while the dashed black line shows the future projections from XENONnT. These limits are extrapolated to tens of TeV masses, assuming an inversely proportional scaling of the local DM abundance with its mass. The model demonstrates its ability to exceed strong upper bounds due to high saturation masses, which is attributed to the effectiveness of coannihilations. The correlation with the Higgs portal coupling values is evident, as the elastic process is mediated by Higgs exchange in the t-channel.}
    \label{DD}
\end{figure}
This situation is explicitly shown in Fig.~\ref{DD}, where we retain the points that saturate the DM budget with an accuracy of about $ 10 \%$, and we plot them alongside the upper bounds reported by DD collaborations.
In this plot, we also verify the explicit dependence of the tree-level DD signal on $|\lambda_{HV}|^2$.
This insensitivity is expected due to the saturation mass being at a few dozen TeV: in this kinematic regime, the nucleon-WIMP cross-section sensitivity evolves inversely with the DM mass.
Therefore, the model is not expected to be probed at DD facilities when the total DM budget consists only of $V^0$.

It is worth noting that, typically in these minimal DM constructions, there is a non-negligible contribution to the DM-nucleon elastic cross-section arising from loop diagrams.
In models with fermionic multiplets, a renormalizable Higgs portal interaction cannot be written, making the loop contribution the only contribution to DD signals \cite{Cirelli:2005uq}.
In our case, however, we expect these contributions to be relatively smaller than the tree-level Higgs interaction we are considering, due to the enormous mass regime required by relic density considerations, which will suppress the signals by means of internal propagators.
Therefore, in this work, we compute only the tree-level contribution.
\begin{table}[t]
    \centering
    \begin{tabular}{|lc|}
        \hline
        \multicolumn{1}{|c|}{\textbf{Parameter}}       & \multicolumn{1}{c|}{\textbf{Value}} \\ \hline
        \multicolumn{2}{|c|}{\textbf{Differential evolution}}                         \\  \hline
        \multicolumn{1}{|c|}{Population size}         & $10$                       \\ \hline
        \multicolumn{1}{|c|}{Tolerance}             & $0.01$                      \\ \hline
        \multicolumn{1}{|c|}{Mutation}        & $[0.7,1.99999]$            \\ \hline
        \multicolumn{1}{|c|}{Recombination}   & $0.15$                     \\ \hline
        \multicolumn{2}{|c|}{\textbf{Lagrangian density}}                             \\ \hline
        \multicolumn{1}{|c|}{$1/\xi$}         & $[-10,-1.5]$               \\ \hline
        \multicolumn{1}{|c|}{$\lambda_{HV}$}  & $[-1,1]$                   \\ \hline
        \multicolumn{1}{|c|}{$M_V$ } & $[10,100]$ TeV                 \\ \hline
    \end{tabular}
    \caption{Specification of the parameters for the random scan, both for the differential evolution method and for the inspected space of the Lagrangian constants.}
    \label{table1}
\end{table}

\subsection{Indirect Detection and Sommerfeld Enhancement}
\label{subsec: Indirect}
In the previous section, we discussed the challenges the model faces in providing observable direct detection features.
To identify observable signatures of this scenario,  we now calculate the indirect signals of $\gamma$ rays produced from DM pair annihilation in specific astrophysical environments, particularly focusing on the center of the Milky Way.
We include the non-relativistic effect known as the Sommerfeld enhancement, which accounts for the resummation of interactions between incident states, potentially leading to a significant increase in the relevant cross-sections for this search strategy \cite{Hisano:2004ds}.

To this end, we solve the Schrödinger equation in the low-velocity limit of the theory given by
\begin{equation}\label{Schrodinger}
    \left( -\frac{1}{M_V} \frac{d^2}{dr^2}-\frac{l(l+1)}{M_V r^2} - E + V(r) -\frac{i \Gamma N}{8 \pi r^2} \delta(r) \right) g_{ab}(r)=0,
\end{equation}
where $a$ characterizes the initial two-particle state and $b$ the final two-particle state.
We are interested in neutral two-particle states when calculating DM annihilation. $M_V$ denotes the tree-level mass of the new vector bosons, $r$ represents the relative distance between the particles, and $E$ denotes the system's kinetic energy, defined as $E=\frac{M_V v^2}{4}$, where $v$ is the relative velocity.
The potential $V(r)$ arises from trilinear interactions between the new vector bosons and the electroweak bosons, including the Higgs, so we write $V(r)=V_{\text{gauge}}(r) + V_{\text{Higgs}}(r)$, comprising Coulomb and Yukawa interactions.
An imaginary term proportional to $\Gamma=\Gamma_{\text{gauge}}+\Gamma_{\text{Higgs}}$, known as the absorptive potential, arises from the low-energy limit of quartic interactions.
$N$ quantifies the number of spin degrees of freedom according to $N=(2l+1)^2$, $\delta(r)$ denotes the Dirac delta function, and $g_{ab}(r)$ represents the wave function characterizing the process $a \rightarrow b$. For explicit calculation of these terms, refer to Appendix~\ref{sec: Appendix Sommerfeld}.

Although polarized initial states should be considered in the general case, the dominant contribution  in the low-velocity regime arises from the $s$-wave, corresponding to $l=0$. 
Hence, we focus solely on cases where the total angular momentum is given by the spin angular momentum.
Thus, we omit the second term in equation \eqref{Schrodinger}.

The boundary conditions for this problem are prescribed as follows \cite{Acevedo:2024ava}:
\begin{equation}
    g_{ab}(r \rightarrow 0) = \delta_{ab},
\end{equation}
and
\begin{equation}
    g_{ab}(r \rightarrow \infty) = d_{ab}(E)e^{i\sqrt{M_V E}r}\exp\left(i\frac{M_V \alpha}{2\sqrt{M_V E}}\log{(2\sqrt{M_V E}r)}\right),
\end{equation}
where $\delta_{ab}$ is the Kronecker delta, $\alpha$ is the fine structure constant, and $d_{ab}$ are known as Sommerfeld factors.
The asymptotic boundary condition for large distances is applicable due to the presence of Coulomb and Yukawa potentials, supported by the fact that the de Broglie wavelength of the exotic bosons is smaller than that of the massive bosons in the SM.
The Sommerfeld factors are computed upon solving equation \eqref{Schrodinger} using the Green's function method, specifically employing the Schwinger-Dyson equations for the system.
Once determined, the optical theorem is utilized, relating the imaginary part of the Green's functions to the thermally averaged total DM annihilation cross-section:
\begin{equation}
    \label{cross2}
    \braket{\sigma v}_{XX^{\prime}}= c\sum_{a,b}\sum_{J,J_{z}}(\Gamma_{XX^{\prime}}^{J,J_{z}})_{ab}~d_{2 a}(E)d_{2 b}^{\ast}(E),
\end{equation}
where $c$ is a normalization constant (2 for $a=b$, 1 for $a\neq b$).
The index fixed at 2 relates to the definition of the vector $\vec{s}$ (see Appendix~\ref{sec: Appendix Sommerfeld}), used to denote DM annihilation processes $(V_{0}V_{0} \longrightarrow XX^{\prime})$ involving SM particles $X$ and $X'$. 

To compute the relevant total and partial annihilation cross-sections, we employed \verb|Mathematica| to solve the Schrödinger equation and determine the Sommerfeld factors. We considered two distinct non-relativistic velocity regimes: $v=10^{-3}$ for assessing indirect signals from galactic DM halos, and  $v=0.3$ to examine the subdominant impact of the Sommerfeld enhancement factor on the relic density calculation.

\begin{figure}[t]
    \centering
    \includegraphics[width=0.7\textwidth]{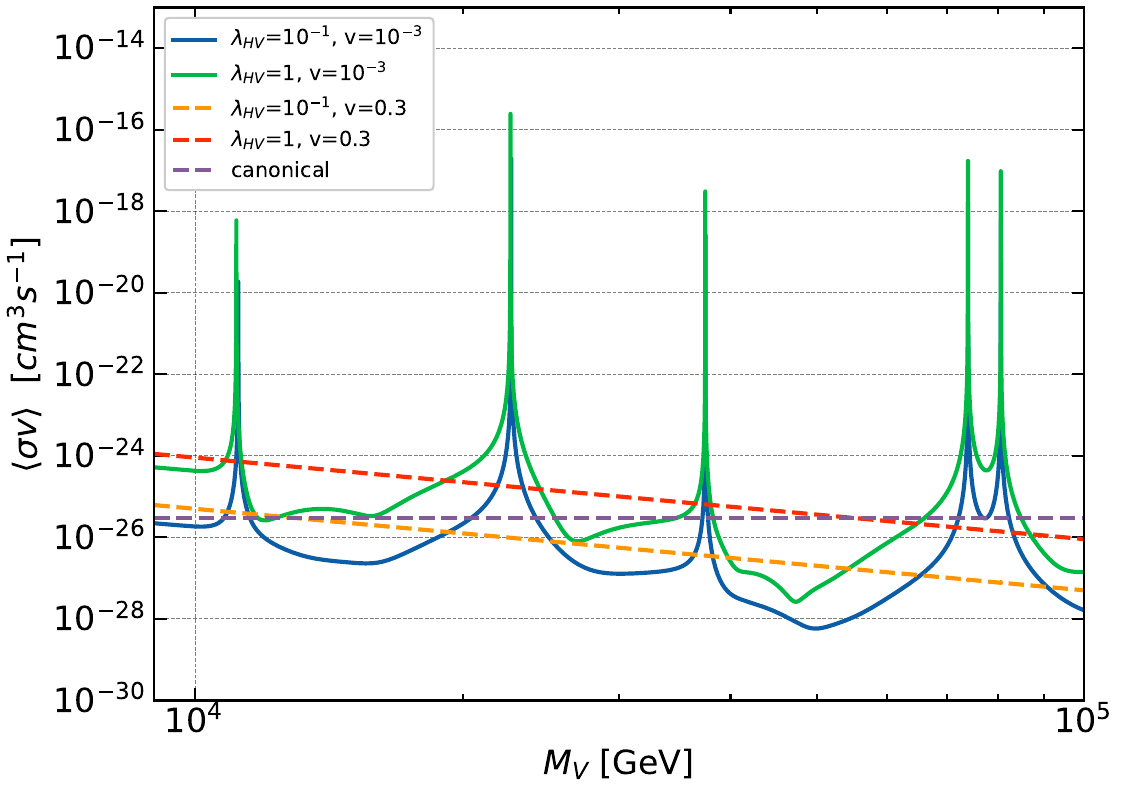}
    \caption{Thermally averaged total annihilation cross-section of DM, considering different values of $\lambda_{HV}$ and relative velocities of incident particles. The purple dashed line represents the canonical annihilation cross-section of $3 \times 10^{-26}$ cm$^3$ s$^{-1}$ for comparison. The green and blue lines exhibit a peak structure at relatively low velocities, contrasting with the dashed red and orange lines where the Sommerfeld enhancement diminishes at higher velocities.}
    \label{Somm1}
\end{figure}
In Fig.~\ref{Somm1} we present the total annihilation cross-section of DM as a function of the tree-level multiplet mass.
As expected, the Sommerfeld factor exhibits a minimal effect on this cross-section for velocities typical during the decoupling era (red and orange lines), when compared to the cross-section magnitudes typical in WIMP scenarios (purple dashed line).
Conversely, at lower velocities, the Sommerfeld effect becomes significant, leading to prominent peaks at masses of $11.2$, $23.1$, $37.5$, $75.03$, and $81.3$ TeV.
This suggests a substantial increase in cosmic ray fluxes originating from DM annihilation and subsequent decays into two-body final states, making current and planned experiments in extraterrestrial particle detection potentially sensitive to this scenario.
Specifically, we focus on $\gamma$ rays as indirect probes of annihilating DM, although it has been shown that antiproton fluxes are particularly important for Higgs portal interactions.

The differential photon flux from DM annihilation represents the number of photons emitted per unit area, time, energy, and solid angle $\Omega$ and is given by
\begin{equation}
    \label{diffphotons}
    \dfrac{d\Phi}{dE_{\gamma}}=\sum_{i}\dfrac{1}{8\pi}J(\Omega)\dfrac{\braket{\sigma v}_{i}}{M_{V}^{2}}\left(\dfrac{dN_{i}}{dE_{\gamma}} \right),    
\end{equation}
where $J(\Omega)$ is a factor dependent on the DM distribution in the region of interest, defined as
\begin{equation}
    \label{Jastro}
    J(\Omega)=\dfrac{1}{\Delta\Omega}\int d\Omega\int d\ell~ \rho(r)^{2}.    
\end{equation}
Here, $\rho(r)$ is the density of DM.
The $J$-factors for various astrophysical regions are approximately $J\approx 10^{19}$ GeV$^2$cm$^{-5}$ for dwarf galaxies, $J\approx 10^{20}$ GeV$^2$cm$^{-5}$  for the Andromeda galaxy, and $J\approx 10^{25}$ GeV$^2$cm$^{-5}$ for the Milky Way's center \cite{Profumo:2013yn}.

On the other hand, $\braket{\sigma v}_{i}$ is the $i$-th partial annihilation cross-section of DM into photons.
For subsequent decays, we employ the narrow-width approximation, multiplying by the branching ratio of final-state SM particles decaying into photons.
$\left(\frac{dN_{i}}{dE_{\gamma}}\right)$ represents the number of photons with energy $E_\gamma$ produced in each event.

\begin{figure}[t]
    \centering
    \subfloat[$V_{0}V_{0}\longrightarrow\gamma\gamma$]{
    \label{flux0}
    \includegraphics[width=0.48\textwidth]{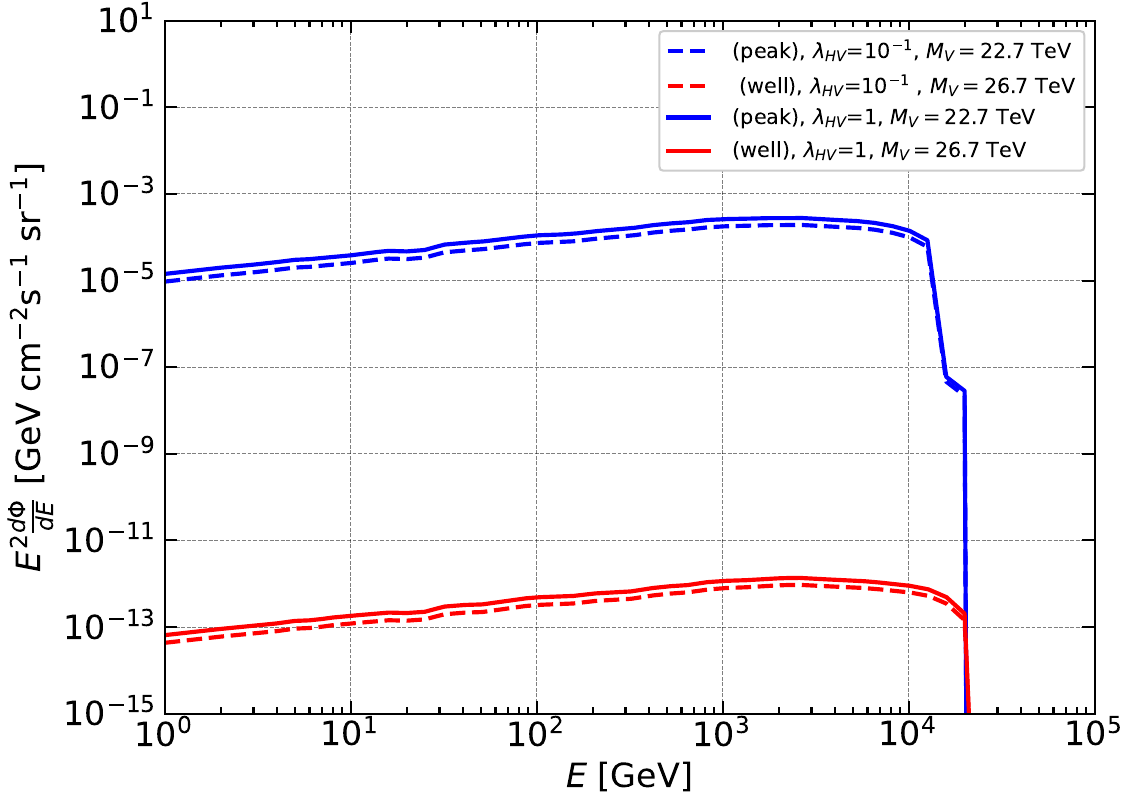}}
    \subfloat[$V_{0}V_{0}\longrightarrow W^{+}W^{-}$]{
    \label{flux0.5}
    \includegraphics[width=0.48\textwidth]{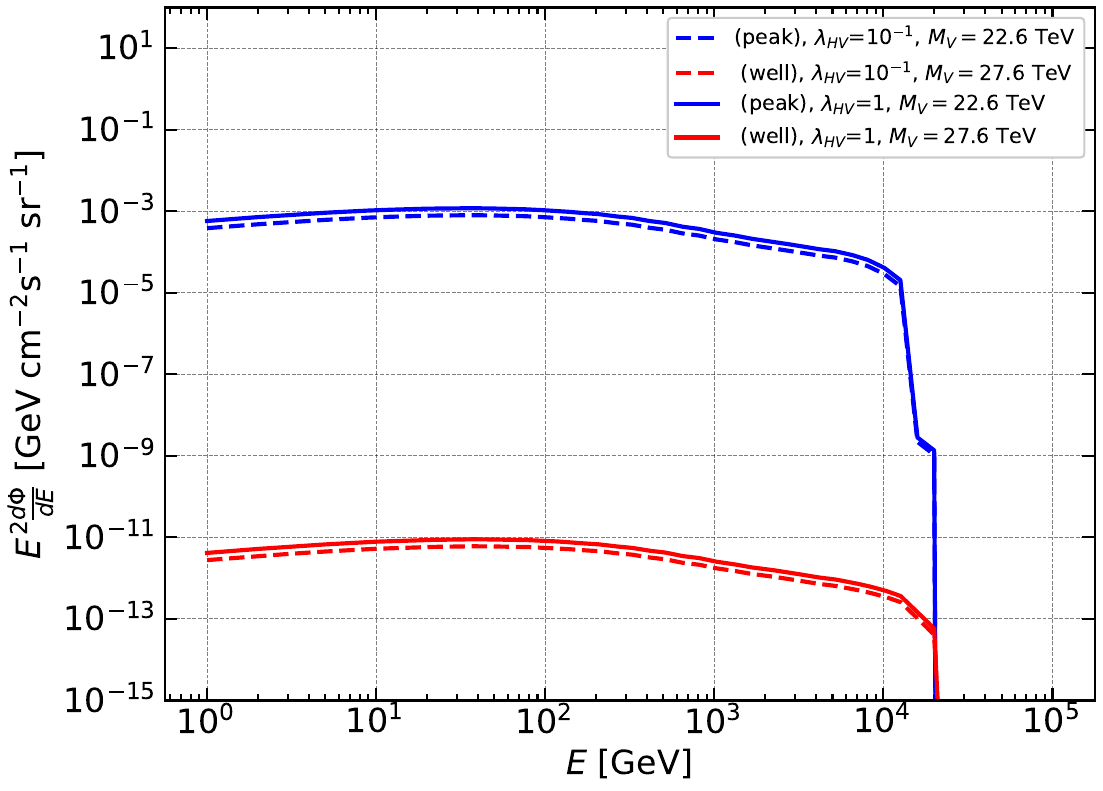}}\\
    \subfloat[$V_{0}V_{0}\longrightarrow ZZ$]{
    \label{flux2}
    \includegraphics[width=0.48\textwidth]{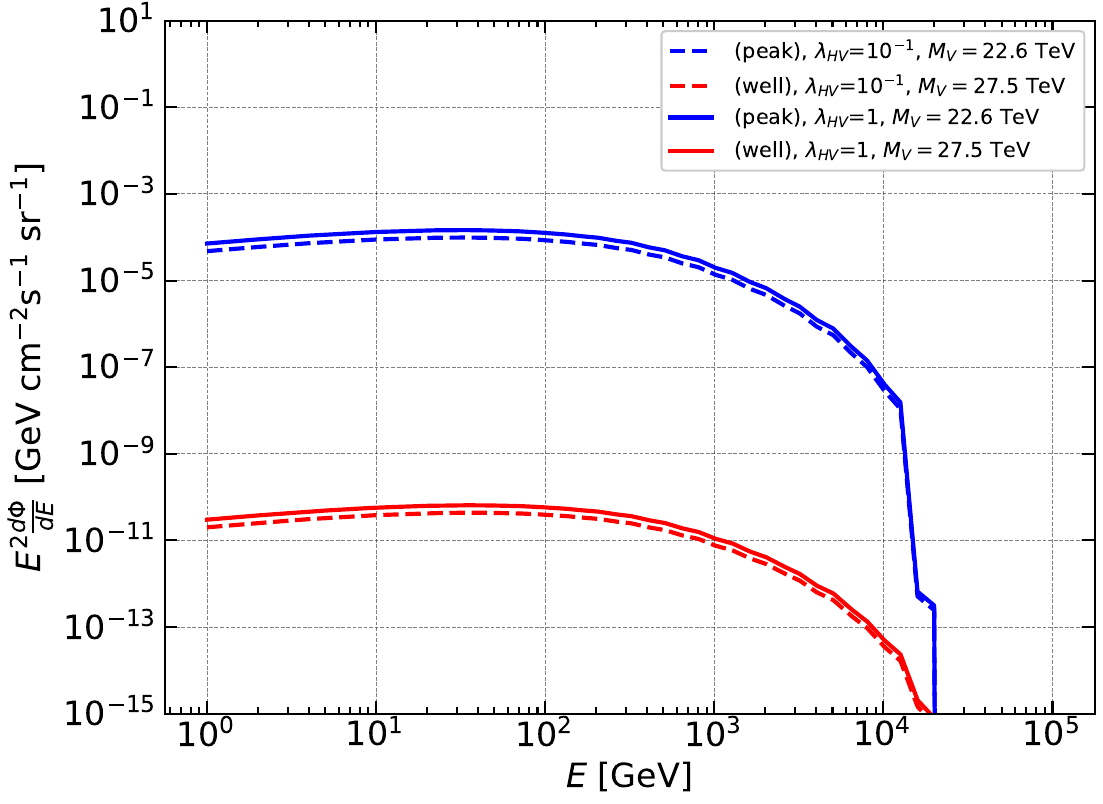}}
    \subfloat[$V_{0}V_{0}\longrightarrow hh$]{
    \label{flux4pi}
    \includegraphics[width=0.48\textwidth]{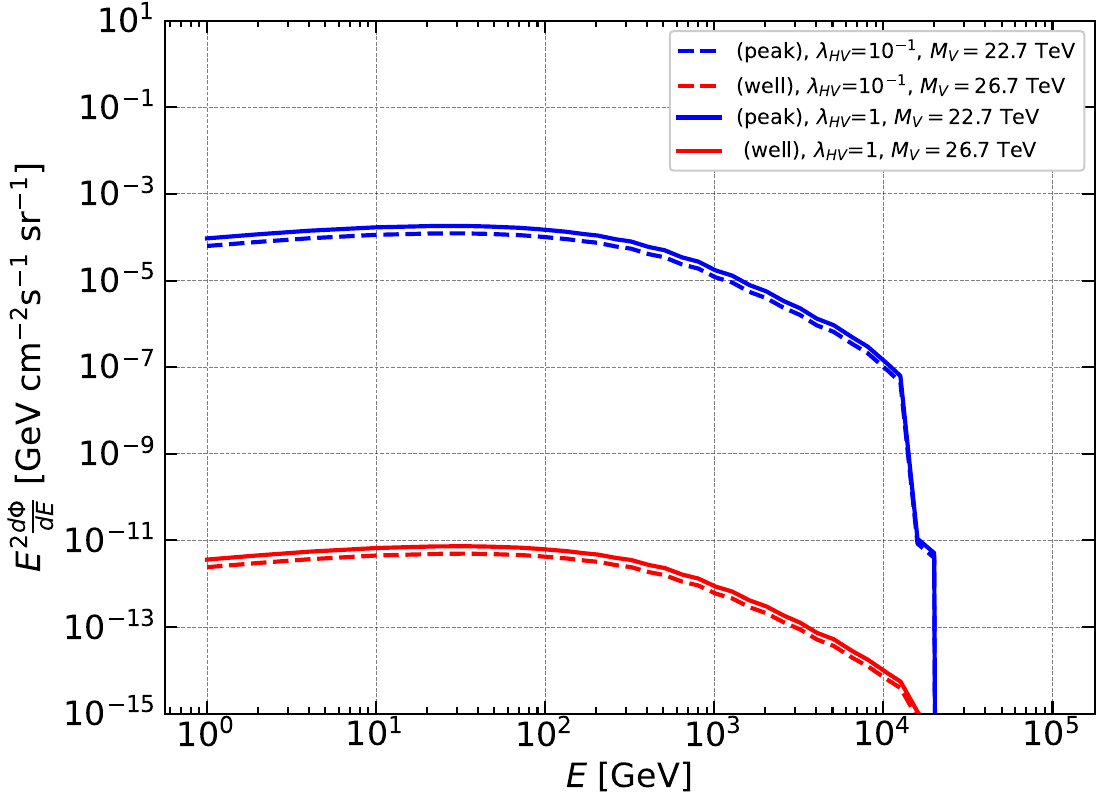}}\\
    \caption{Photon flux resulting from the annihilation of DM particles, assuming the NFW profile. We considered a region of the center of the Milky Way with solid angle $\Delta\Omega=0.96\times10^{-3}$ sr (see Table 2 in \cite{Cirelli:2010xx}). We fix $\xi^{-1}=-1.5$.}
    \label{Flux}
\end{figure}
We calculated the differential photon flux using \verb|Mathematica|, following the methodology outlined in \cite{Cirelli:2010xx}.
In Fig.~\ref{Flux} we depict the photon flux for various annihilation channels, $\gamma\gamma$, $W^{+}W^{-}$, $ZZ$, and $hh$. 
We compare two characteristic cases of DM mass, one predicting a peak in the Sommerfeld factor and one located at a well in the cross-section as a function of mass.
The plots reveal a significant enhancement in the former scenario, with the most prominent flux originating from annihilation into charged weak bosons and Higgs particles.
\begin{figure}[t]
    \centering
    \includegraphics[width=0.7\textwidth]{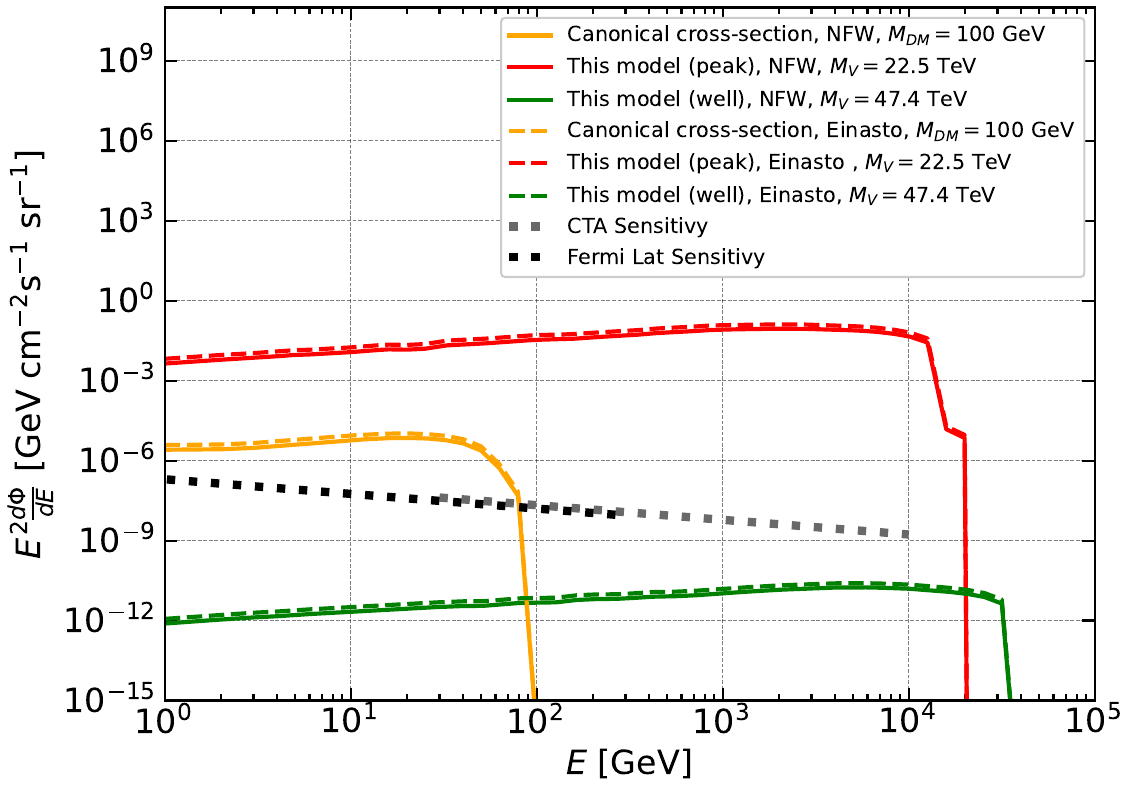}
    \caption{Total photon flux from DM annihilation, for Einasto and NFW profiles, considering a region of the center of the Milky Way with solid angle $\Delta\Omega=0.96\times10^{-3}$ sr (see Table 2 in \cite{Cirelli:2010xx}). For comparison, we include the sensitivity of Fermi-LAT \cite{Ibarra:2012dw} and CTA \cite{Silverwood:2014yza}, as well as the flux given by a typical WIMP with mass of $100$ GeV and canonical annihilation cross-section. We fixed the parameters as $\xi^{-1}=-1.5$ and $\lambda = 1$.}
    \label{totalflux}
\end{figure}
In Fig.~\ref{totalflux} we present the total photon flux for indirect detection of DM, comparing with a typical WIMP scenario and the sensitivity of the Fermi-LAT and CTA experiments.
These results highlight that non-perturbative effects render certain regions of the parameter space accessible to current and planned photon detectors.
This is particularly relevant given the lower direct detection rates calculated in previous sections.

\subsection{Collider Phenomenology}
\label{subsec: Collider}
In this section, we consider some signatures of the model in collider facilities.
As stated before, the effectiveness of coahinillations pushes the DM mass required to satisfy the relic density constraint to tens of TeV, depending on the choice of couplings, with the $\xi$ parameter being particularly significant.
Therefore, we will focus on the prospects of signals in the Future Circular Collider (FCC) \cite{FCC:2018vvp}, with an expected center of mass energy of $\sqrt{s}=100$ TeV, instead of the present-day Large Hadron Collider, which is capable of producing lighter particles, resulting in an under abundant DM budget in this scenario.

In the following subsections, we compute the decay widths and mean decay lengths of the charged components of the exotic vector multiplet.
For concreteness, we consider the mass splitting given in Fig.~\ref{splitting} for the case $Q=M_V$.
Notice that masses of hundreds of GeV allow a mass splitting large enough to open mesonic decay channels. We then present the results of a Monte-Carlo simulation of proton-proton collisions with a beam energy of $50$ TeV, studying the $VVh$ production process in order to quantify the influence of the Lagrangian parameters $\xi$ and $\lambda_{HV}$ in this search strategy. 

\subsubsection{Charged Vector Decays}
The mass splitting shown in Sec.~\ref{sec: The mass splitting} between these particles is responsible for the instability of the exotic charged bosons, making $V^0$ the lightest and a good DM candidate.
On one hand, the double-charged boson can decay into the neutral one or the singly charged component, accompanied by hadrons, leptons, or photons, depending on the kinematics.
Meanwhile, $V^\pm$ can only decay into the DM candidate, accompanied by leptons or a hadron. 

The most relevant decay channels of $V^+$ due to mass splitting are $V^+ \to V^0 l^+\nu_l$ (with $l=\mu,e$), and $V^+\to V^0 u\Bar{d}$.
For the double-charged boson $V^{++}$ the decay modes are analogous: $V^{++} \to V^+ l^+\nu_l$ and $V^{++}\to V^+ u\Bar{d}$.
Additionally, $V^{++}$ can decay into final states that involve heavier quarks. 

In addition, four-body decay channels are also available for charged components: $V^{+}\to V^0 q_i\Bar{q}_j\gamma$, $V^{++}\to V^+q_i\Bar{q}_j\gamma$ and $ V^{++}\to V^+ q_i\Bar{q}_jf\Bar{f}$, where an intermediate off-shell $W$ or $Z$ boson propagator mediates the process.
The quark flavors in the final states depend on the mass splitting.
However, these four-body decay channels are negligible for any parameter configuration due to phase space suppression, and therefore they are not considered further in this analysis.

Nevertheless, since the mass splitting allows for mesonic final states, accurate calculations of partial widths require consideration of QCD corrections \cite{Belyaev:2018xpf,Belyaev:2016lok}.
To accurately handle these processes, we employ the following effective Lagrangian for $W-X$ mixing where $X$ stands for a meson:
\begin{equation}
    \mathcal{L}_{WX}=\sum_{X^-=\pi^- , K^- }\dfrac{gf_X V_{qq'}}{2\sqrt{2}}W_\mu^+\partial^\mu X^-   \,- \sum_{X^+ =\rho^+ } \dfrac{gm_X^2  V_{qq'}w_X}{2g_\rho}W_\mu^-X^\mu  + h.c.,
    \label{LwX}
\end{equation}
where $V_{qq'}$ represents elements of the CKM matrix, $f_X$ and $w_X$ are related to the meson decay constant, $m_X$ the mass of the meson and $g_\rho$ is the $\rho \pi\pi$ constant\footnote{We use $f_\pi=0.130$ GeV , $f_K=0.156$ GeV and $w_\rho=1$ \cite{Lichard:1997ya}.}. 
We restrict our consideration to mesons with masses below approximately 0.8 GeV.
Therefore, this effective Lagrangian includes both pseudoscalar and vector mesons.
\begin{figure}[t]
    \centering
    \includegraphics[width=0.35\textwidth]{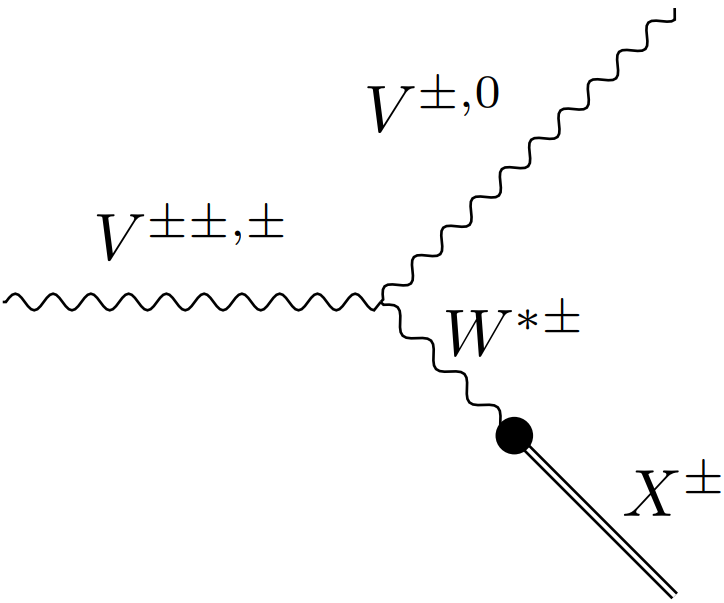}
    \includegraphics[width=0.35\textwidth]{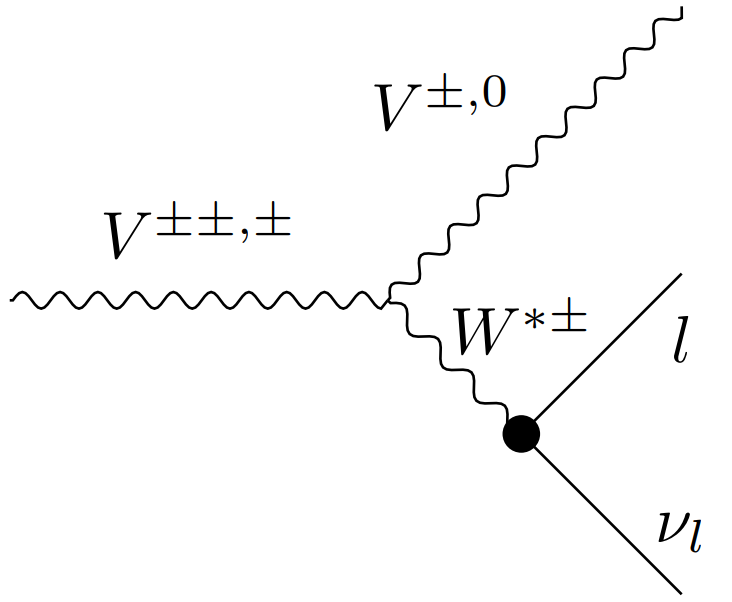}
    \caption{Principal decay channels of the charged components. Left: Hadronic channel via effective $W-X$ mixing leading to a final state of two bodies, where $X$ represents the mesons allowed by the mass differences. Right: Leptonic channel where $l$ denotes $e$ or $\mu$.}
    \label{decays_figures}
\end{figure}

The effective Lagrangian \eqref{LwX} describes the processes illustrated by the Feynman diagram on the left of Fig.~\ref{decays_figures}.
The explicit forms of the decay widths into pseudoscalar and vectorial mesonic final states are
\begin{align}
    \begin{split}
        \Gamma_{(esc)} = &\frac{ g^4 f_X^2 V_{qq'}^2 \Delta_{+}^2(2 M_{V^+}-\Delta_{+})^2 \sqrt{(\Delta_{+}^2-m_X^2) ((2 M_{V^+}-\Delta_{+})^2-m_X^2)}}{512 \pi  M_{V^+}^5 M_W^4 (M_{V^+}-\Delta_{+})^2} \\
        &\times \left(2 (M_{V^+}-\Delta_{+})^2 (5 M_{V^+}^2-m_X^2)+(M_{V^+}^2-m_X^2)^2+(M_{V^+}-\Delta_{+})^4\right),
    \end{split}\\[0.4cm]
    \begin{split}
        \Gamma_{(vec)} = &\frac{g^4 m_X^2 w_X^2 V_{qq'}^2 ((\Delta_{+}^2-m_X^2) ((2 M_{V^+}-\Delta_{+})^2-m_X^2))^{3/2}}{256 \pi  g_{\rho }^2 M_{V^+}^5 M_W^4 (M_{V^+}-\Delta_{+})^2} \\
        &\times \left(2 (M_{V^+}-\Delta_{+})^2 (5 M_{V^+}^2+m_X^2)+(M_{V^+}-\Delta_{+})^4+M_{V^+}^4\right). \\
    \end{split}
\end{align}
For the doubly charged component $V^{++}$, the expressions are analogous, changing $\Delta_{+} \to \Delta_{++}$, $M_{V^+}\to M_{V^{++}}$ and a factor $2/3$ due to the $VVW$ vertex. 
The Feynman diagram on the right of Fig.~\ref{decays_figures}, involving leptonic final states, was calculated numerically using \verb|CalcHEP| \cite{Belyaev:2012qa}.
\begin{figure}[t]
    \centering
    \begin{subfigure}{0.4965\textwidth}
        \includegraphics[width=\textwidth]{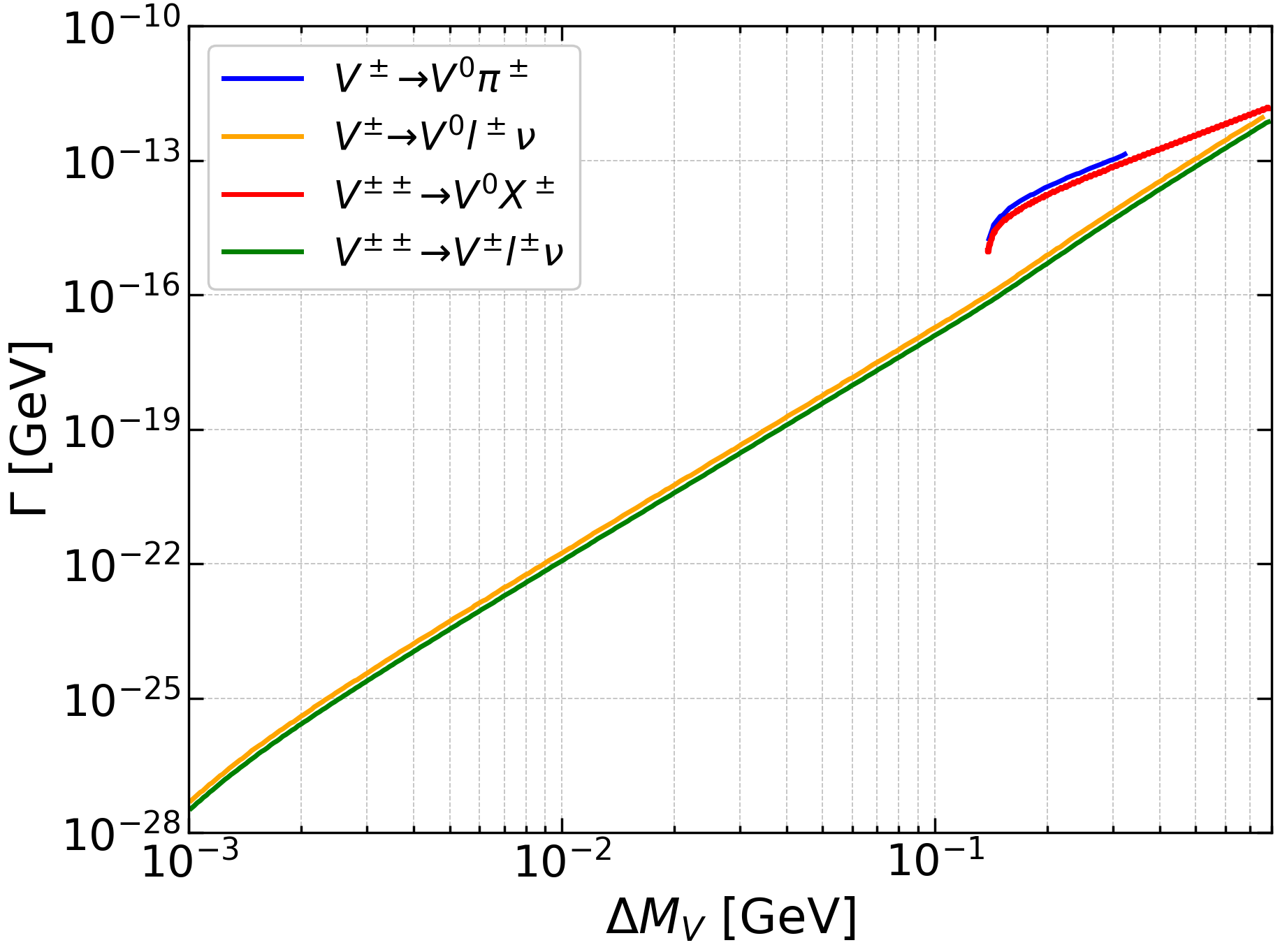}
    \end{subfigure}
    \begin{subfigure}{0.4965\textwidth}
        \includegraphics[width=\textwidth]{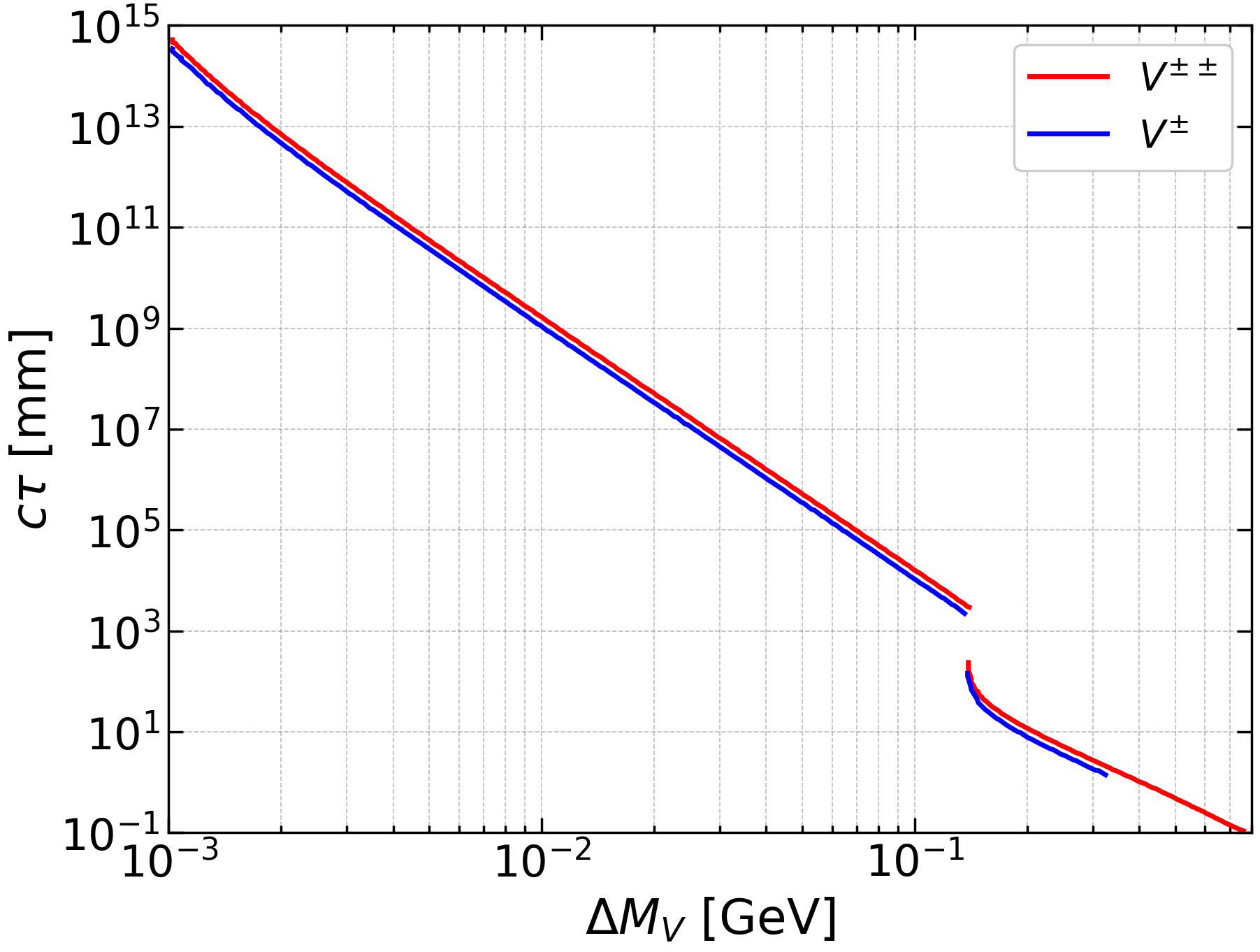}
    \end{subfigure}
    \caption{Left: Partial decay widths of the charged vector components as a function of the mass splitting between the mother particle and its partner, differing by one unit of charge. The red and blue lines indicate hadronic decay channels, while the orange and green lines depict the leptonic channels. Right: Mean decay length of the charged vector components in mm. Red and blue lines correspond to $V^{++}$ and $V^{\pm}$ respectively. The plots are computed fixing $\xi^{-1}=-5$.}
    \label{widths_large}
\end{figure}

The results of these calculations are presented in Fig.~\ref{widths_large}, where we show the different contributions to the partial decay widths as a function of the mass splitting on the left and the corresponding mean decay length of the two charged components on the right. We observe a significant contribution from the leptonic channel to the total width.
As the mass splitting increases, so does the total width.
When $\Delta M \gtrsim 140$ MeV the hadronic channel becomes open, significantly increasing the decay width and becoming more relevant than the leptonic channel.
Hence, for 140 MeV $\lesssim \Delta M \lesssim$ 400 MeV, the leptonic channel can be ignored, with the $V^{++,+} \to V^{+,0} \pi^+ $ channel being the most relevant decay mode.
However, beyond this range, both channels must be considered.
The production of kaons and rho mesons as final states is only accessible in the $V^{++}$ decay, leading to a decay width of $\Gamma \sim 10^{-11}$ GeV.
Importantly, the partial width decay is independent of the $\xi$ term, and the mass of the charged vector does not significantly influence the width once the mass splitting is fixed.

Furthermore, we observe that a mass splitting exceeding the pion mass results in a significant reduction of the mean decay length.
When the mesonic channel opens, the $c\tau$ value decreases sharply, by about an order of magnitude.
A mass difference $\Delta M \gtrsim$ 400 MeV produces a mean length $l =c\tau \lesssim$ 1 mm, leading to a challenging disappearing track to detect within the tracker, thus motivating the study of these exotic particles as long-lived particles.
Conversely, for $\Delta M \lesssim$ 150 MeV, particularly where the leptonic channel dominates, and especially for $\Delta M \lesssim 50$ MeV, the charged bosons exhibit a large mean decay length (see, e.g., \cite{Avila:2021mwg,Belyaev:2020wok,R:2020odv}).

\subsubsection{VVh Production}
We conducted simulations for $VVh$ production in proton collisions using the \verb|CalcHEP| package, incorporating the one-loop radiative correction for $hgg$ effective interactions available from the HEP Model Database \cite{Belyaev:2020lon}.
The environment considered is the FCC-hh featuring a center-of-mass energy of $\sqrt{s}=100$ TeV, and a Higgs transverse momentum greater than $0.2$ TeV, with \verb|cteq6l(proton)| as the parton distribution function \cite{Pumplin:2002vw}. 
It is important to note that we did not account for the subsequent decays of the charged components into the neutral one through the chain decay $V^{++}\to V^+ \to V^0$. This omission could affect the detectability of these particles, especially when the corresponding mean decay length is shorter than the distance from the collision point to the detectors.

We have verified that the mass difference does not have a significant impact in the cross sections, so we fixed it at a general value of $\Delta_{+}$= 0.2 GeV.
Once we inferred the cross-sections from the simulation, we considered the Higgs decay into two $b$ quarks, as this is its primary decay channel, and its actual reconstruction is expected in more accurate estimations.
The branching ratio is estimated to be $Br(h\to b\Bar{b})\approx 0.58$ \cite{ATLAS:2022vkf} and the reconstruction efficiency for this channel is $\epsilon_{eff-b}\approx 0.80$ \cite{ATLAS:2017juw,ATLAS:2021qou,CMS:2012feb}.
This simple initial approach is sufficient to understand some aspects of the phenomenology.
An in-depth analysis can be done considering the efficiencies provided in the references \cite{R:2020odv,Saito:2019rtg,Chiang:2020rcv} for processes such as mono-jet or charged companion production.
Within our considerations, the effective cross-section is given by
\begin{equation}
    \sigma_{eff}(pp\rightarrow VVb\bar{b})= Br(h\rightarrow b\bar{b})\cdot \epsilon_{eff-b}\cdot\displaystyle \sigma_{c}(pp\rightarrow VVh),
\end{equation}
where $\sigma_c(pp\rightarrow VVh)$ corresponds to the numerically inferred cross-section. 

\begin{figure}[t]
   \centering
   \includegraphics[width=0.8\textwidth]{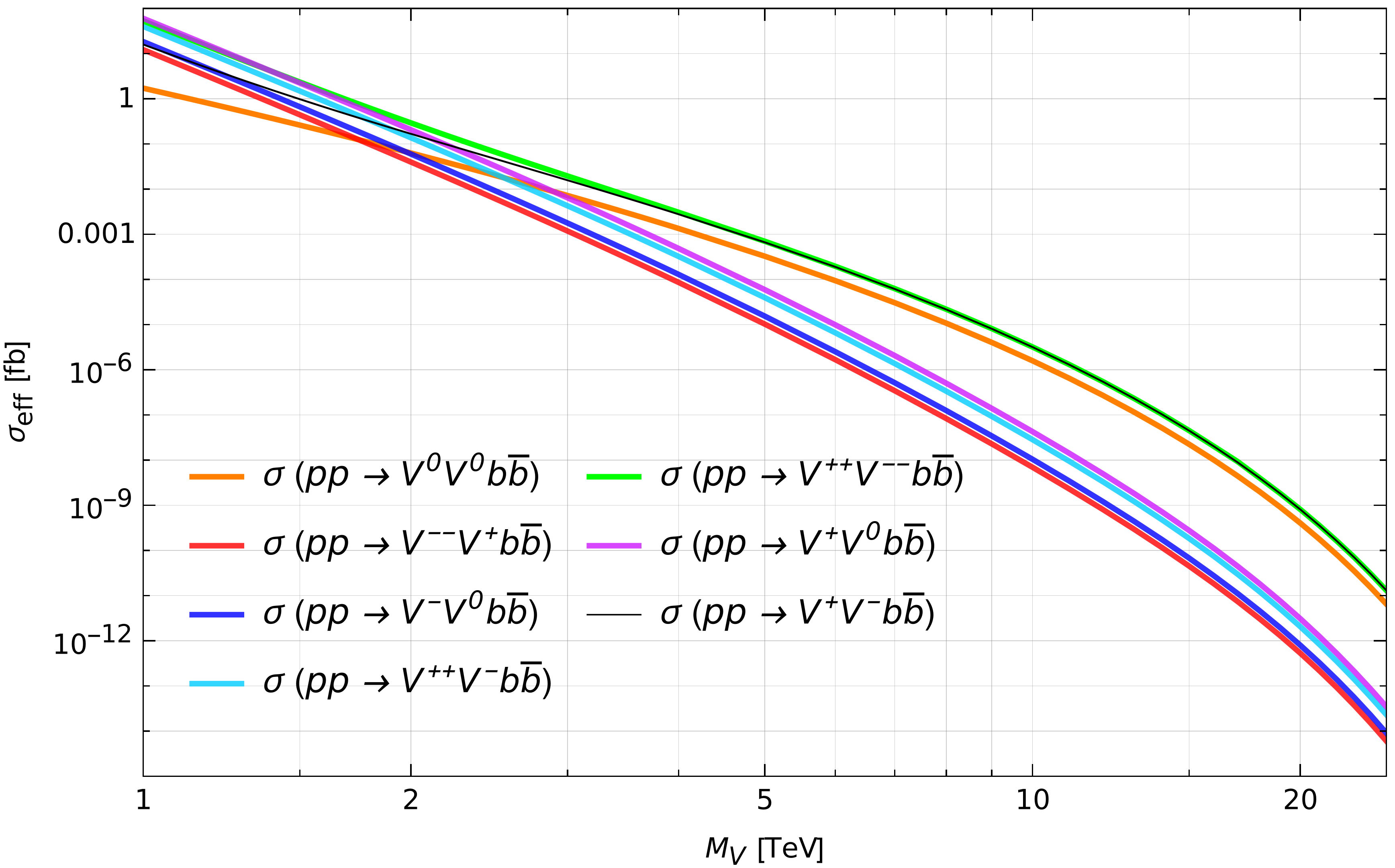}
   \caption{Effective cross-section for the process $pp\rightarrow VVh\to VVb\Bar{b}$ at the FCC. The parameters are constrained by unitarity and fixed as $\lambda_{HV}=1$, $\Delta_{+}=0.2$ GeV, and $\xi^{-1}=-5$. It is observed that for high masses, where the model achieves relic density saturation, the largest cross-sections arise from neutral channels: $VV=V^{++}V^{--},V^{+}V^{-},V^{0}V^{0}.$}
   \label{mh1}
\end{figure}
The results of these calculations are shown in Fig.~\ref{mh1}.
In the first plot, we observe that the cross-section for processes producing a neutral final state $V^0V^0, V^+V^-,$ or $V^{++}V^{--}$ is dominant in the mass region $M_V \gtrsim 2.5$ TeV.
This corresponds to the regime in which the DM candidate can indeed account for the entire DM population inferred from CMB data.
In contrast, when the DM mass lies below these values, the production cross-section of a charged final state becomes comparable.

\begin{figure}[t]
    \centering
    \includegraphics[width=0.49\textwidth]{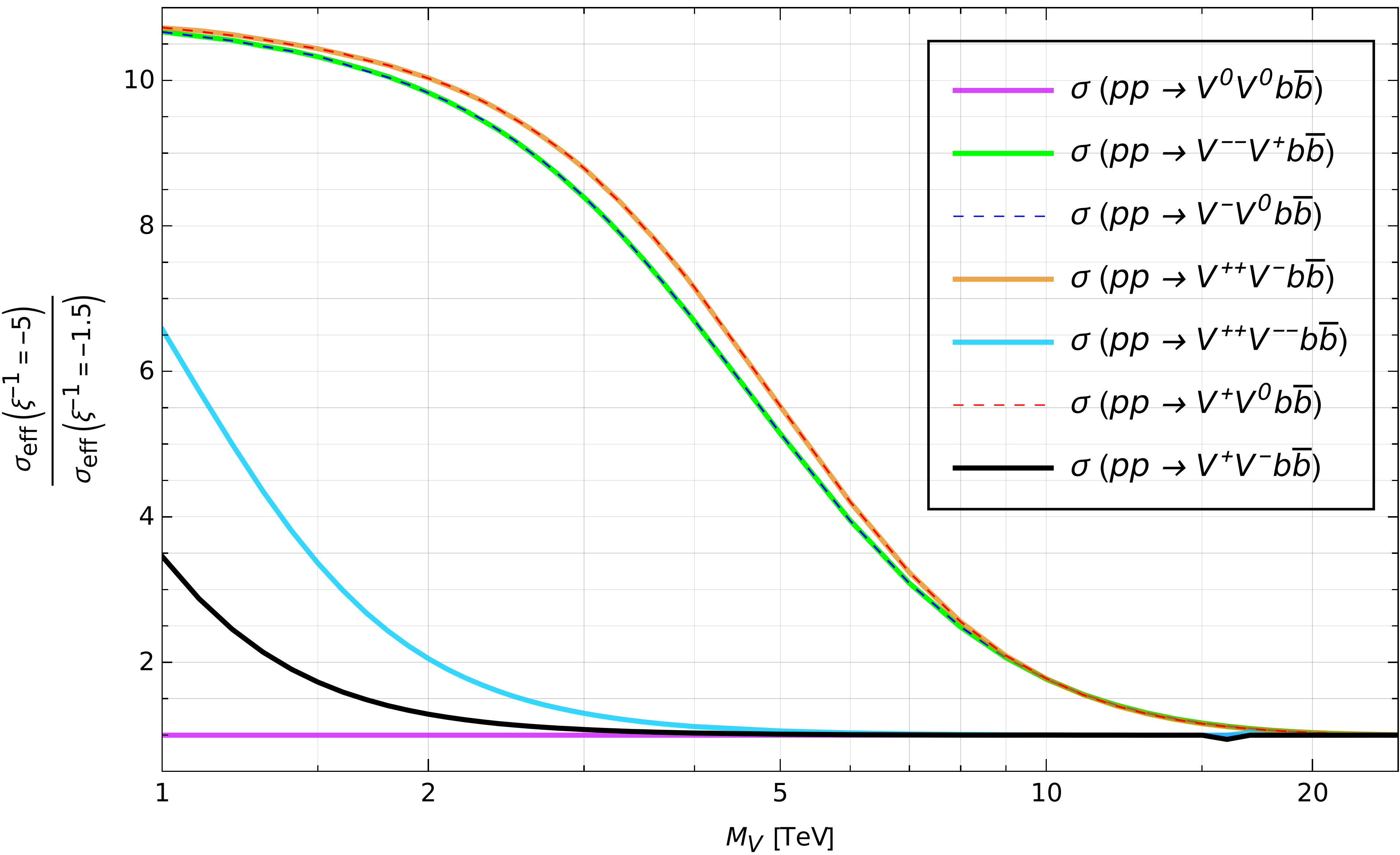}
    \includegraphics[width=0.49\textwidth]{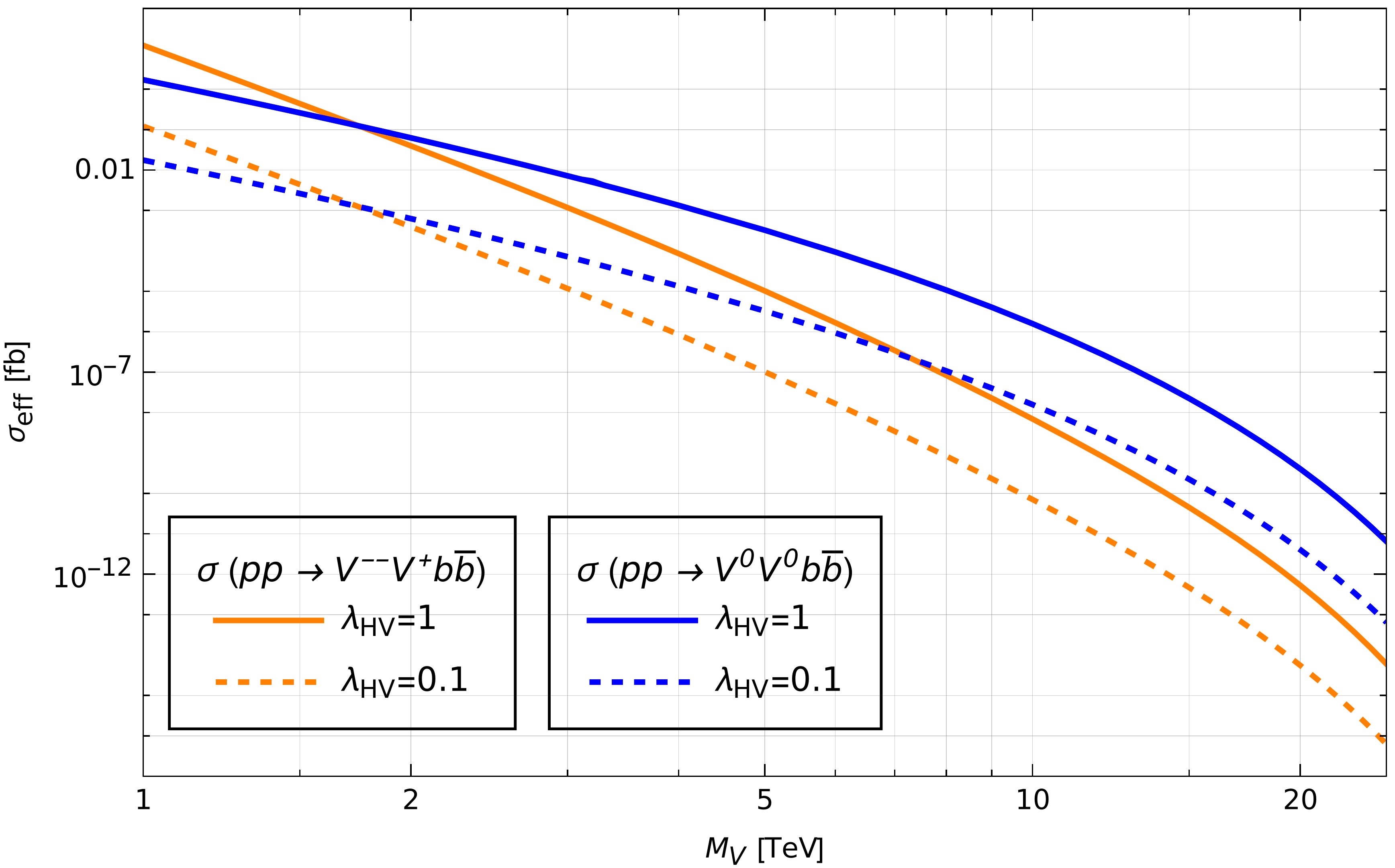}
    \caption{Left: Ratio of effective cross-sections using $\xi^{-1}=-5$ and $\xi^{-1}=-1.5$ with $\lambda_{HV}=1$ fixed. This illustrates the dependence of processes on $\xi^{-1}$ particularly affecting charged channels at low masses, while becoming irrelevant at high masses. Right: Cross-sections for different values of $\lambda_{HV}$ with $\xi^{-1}$ fixed.}
    \label{VVh-production}
\end{figure}
To illustrate the impact of the Lagrangian density coupling constants on the production of a $VVh$ final state, we present some results in Fig.~\ref{VVh-production}. 
In the left plot, we observe the influence of the $\xi$ parameter on the relevant cross-sections.
It primarily affects the production of charged final states, while also having a noticeable but softer effect on neutral final states composed of charged particle-antiparticle pairs.
Additionally, $\xi$ does not impact the production of a pair of DM particles, as the electroweak mediator contribution is negligible compared to gluon fusion.
This is also why the impact of $\xi$ on all processes is reduced for $M_V\gtrsim 20$ TeV.
In the right plot, we see that the Higgs portal coupling constant $\lambda_{HV}$ affects production independently of the tree-level mass of the new degrees of freedom, with its effect manifesting as a quadratic modulation\footnote{We only show two channels for better visualization, the other channels Exhibit the same behavior.}.

In the following we focus on the missing transverse momentum distribution of the Mono-Higgs process, this is the $V^0V^0 h$ production.
We use a benchmark similar to that in previous sections for this analysis, with  $\xi^{-1}=-5$ and $\Delta M= 0.2$ GeV.
Also, we assume a 6\% smearing in the momenta of the final state particles. 

As mentioned, if the charged particles have a sufficiently low $c\tau$, they could contribute to the missing energy in this search strategy. 
One potential method to detect these charged states is through the leptonic decay channel $V^{++,+}\to V^{+,0} l\nu_l$.
However, due to the small mass differences between the new vector states, these leptons would be very soft, significantly reducing their reconstruction efficiency \cite{ATLAS:2019jvq,ATLAS:2020auj}. 
A similar situation occurs in the hadronic decay channel, where the momentum from the meson is also too soft for reliable reconstruction, resulting in a disappearing track signature.
Therefore, we did not consider the contribution to missing transverse momentum from the charged vectors in our analysis.

\begin{figure}[t]
    \centering
    \includegraphics[width=0.8\textwidth]{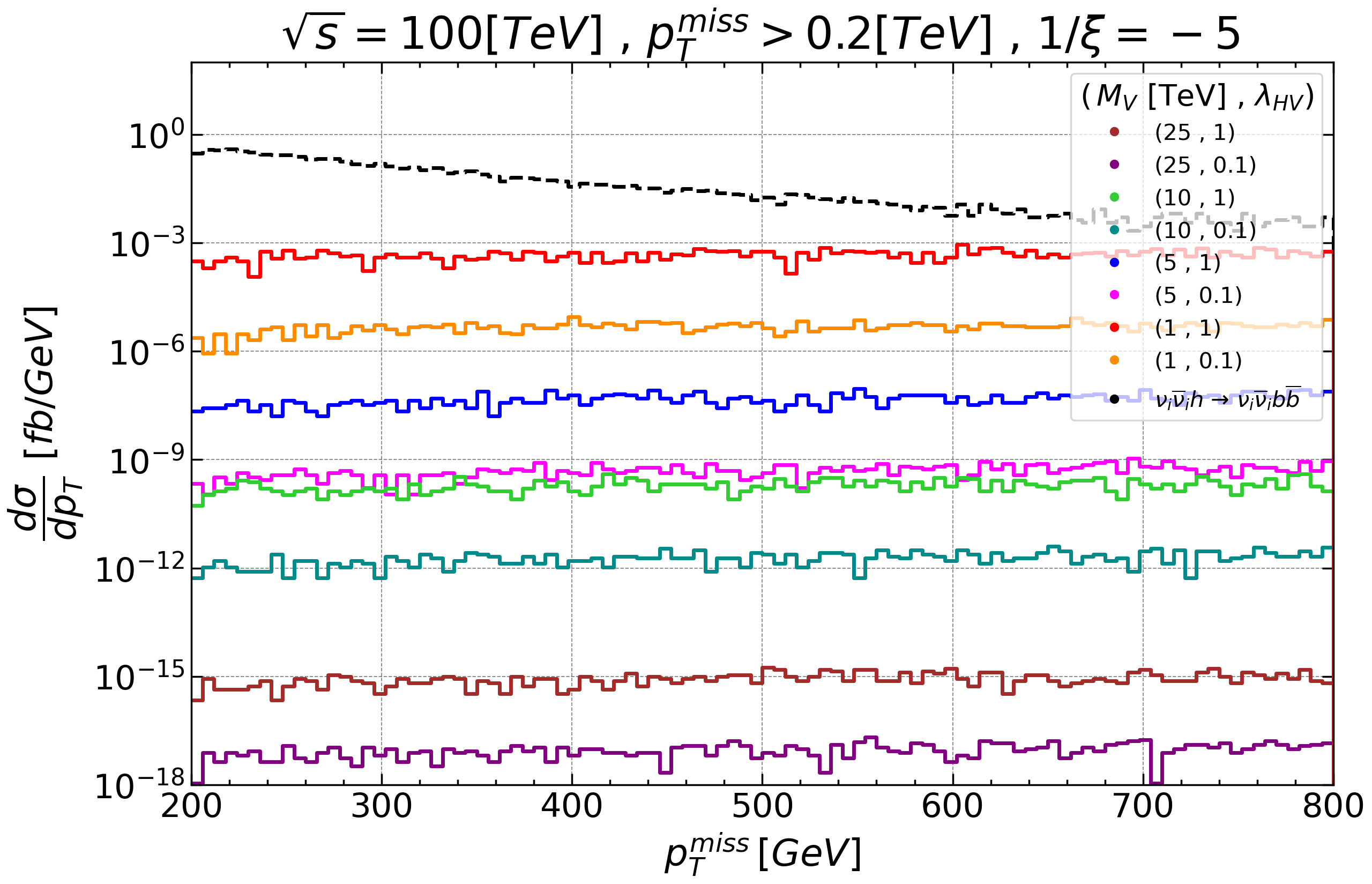}
    \caption{Histogram of the differential cross-section as a function of the missing transverse momentum for different masses of the DM candidate and $\lambda_{HV}$, at $100$ TeV of energy for the $pp\to V^0V^0h\to V^0V^0b\Bar{b}$ process. The black dashed line represents the irreducible background $pp\to Zh\to \nu_i\Bar{\nu}_ib\Bar{b}$.}
    \label{mono-higgs}
\end{figure}
To contrast our scenario with the SM we considered the main irreducible background process $pp\rightarrow Zh\rightarrow\nu\bar{\nu}h$  \cite{Belyaev:2020wok}.
The differential cross-section distribution as a function of missing transverse momentum for various values of $M_V$ and $\lambda_{HV}$ is depicted in Fig.~\ref{mono-higgs}.
From the plot, we observe that the most significant signals appear for relatively small values of $M_V$ and larger  $|\lambda_{HV}|=1$.
The quadratic dependence noted previously in the right plot of Fig.~\ref{VVh-production} is also evident in the momentum distribution.
However, even with masses as low as $1$ TeV, the model does not surpass the irreducible background, leading to a  signal that decreases several orders of magnitude for the mass regime where $V^0$ constitutes the entire DM population.
This renders the signal practically undetectable in these conditions. 

\section{Conclusions}
\label{sec: Conclusions}
In this work, we introduced an extension of the SM featuring an exotic spin-1 boson in the 5-dimensional representation of weak isospin.
Addressing unitarity violation through partial-wave expansion analysis, we constrained the interactions within the spin-1 boson multiplet and its interactions with SM bosons. 
Notably, the calculated mass differences between components of the quintuplet, critical for DM phenomenology, were found to be smaller than those typically induced in other minimal DM models.
Then we studied the implications of non-minimal interactions and the large number of coannihilation channels for DM production, concluding that the DM candidate should have masses of at least $\sim 20$ TeV to satisfy the relic density constraint. 

We studied the phenomenology of the model, exploring direct and indirect search strategies, as well as collider signals at the FCC. 
An exhaustive random scan indicated that this model does not predict a detectable direct detection signal suitable for present- or next-generation instruments. 
Furthermore, utilizing numerical solutions of the non-relativistic Schrödinger equation, we evaluated the Sommerfeld enhancement effect, identifying promising prospects for detecting this scenario through enhanced photon fluxes from the galactic center.
As a first approach to $VVh$ production at the FCC, we conducted a Monte-Carlo simulation to assess $VVh$ production,  revealing challenges in identifying this degrees of freedom even in future experiments.
On the other hand, the mass splitting effect on the mean decay length of the charged components highlights interesting opportunities for exploring long-lived particle scenarios at the FCC.

\newpage
\section*{Acknowledgments}
We would like to thank Jeremy Echeverría, Claudio Dib, Diego Aristizabal, Paola Arias and Bastián Diaz, for useful discussions.
This project has been financed by UTFSM doctoral grants 008/2018 and 003/2020, UTFSM master grants 048/2022, 177/2023 and 091/2024,
PIIC 2021-II and 2022-I, DPP, UTFSM,
ANID-Chile Grants 21210952 and 21210616,
ANID-Chile FONDECYT grant 1230110 and 1170171,
ANID PIA/APOYO AFB230003
and ANID Millenium Science Initiative ANID-ICN 2019-044.

\bibliography{references}
\bibliographystyle{unsrt}

\appendix
\section{Longitudinally Polarized Scattering Amplitudes}
\label{sec: Appendix A}
For the purposes of Sec.~\ref{sec: Perturbative Unitarity}, we present show the expressions for the partial amplitudes of various processes involving the longitudinal components of the newly proposed particles and the massive bosons of the SM.
Although the analysis was conducted on a general gauge for the $W$ and $Z$ bosons, the scattering amplitude remains gauge-independent. Terms that scale as $\mathcal{O}(M_V^2/s)$ or are independent of $s$ are omitted, as our focus here is on high-energy regimes.\\

\footnotesize 
\underline{$V^{++}V^{--}\to V^{++} V^{--}$}
\begin{equation*}
    a_0(s)=\frac{s \left(\left(5 \alpha_1+6 \alpha_2+5 \alpha_3\right) g^2 M_V^2+4 \kappa g^4 \left((4 \kappa+2) M_V^2+3 \kappa M_W^2\right)+\lambda_{HV} ^2 M_W^2\right)}{32 \pi  g^2 M_V^4}-\frac{\left(3 \alpha_1+4 \alpha_2+3 \alpha_3+16 \kappa^2 g^2\right)s^2}{96 \pi  M_V^4}.
\end{equation*}

\underline{$V^{++}V^{--}\to V^{+} V^{-}$}
\begin{equation*}
    a_0(s)=\frac{s \left(-g^4 \kappa  \left((4 \kappa +2) M_V^2+3 \kappa  M_W^2\right)+\left(\alpha _1+2 \alpha _2+\alpha _3\right) g^2 M_V^2+\lambda_{HV} ^2 M_W^2\right)}{16 \pi  g^2 M_V^4}-\frac{ \left(\alpha _1+3 \alpha _2+\alpha _3-8 g^2 \kappa ^2\right)s^2}{96 \pi  M_V^4}.
\end{equation*}

\underline{$V^{++}V^{--}\to V^{0} V^{0}$}
\begin{equation*}
    a_0(s)=\frac{s \left(\alpha_1+2 \alpha_2+\alpha_3+\frac{\lambda_{HV} ^2 M_W^2}{g^2 M_V^2}\right)}{16 \pi  M_V^2}-\frac{\left(\alpha_1+3 \alpha_2+\alpha_3\right) s^2}{96 \pi  M_V^4}.
\end{equation*}

\underline{$V^{+}V^{-}\to V^{0} V^{0}$}
\begin{equation*}
    a_0(s)=\frac{s \left(\alpha _1+2 \alpha _2+\alpha _3+\frac{\lambda_{HV} ^2 M_W^2-3 g^4 \kappa  \left((4 \kappa +2) M_V^2+3 \kappa  M_W^2\right)}{g^2 M_V^2}\right)}{16 \pi  M_V^2}-\frac{\left(\alpha _1+3 \alpha _2+\alpha _3-24 g^2 \kappa ^2\right)s^2 }{96 \pi  M_V^4}.
\end{equation*}

\underline{$V^{+}V^{-}\to V^{+} V^{-}$}
\begin{equation*}
    a_0(s)=\frac{s \left(\left(5 \alpha_1+6 \alpha_2+5 \alpha_3\right) g^2 M_V^2+\kappa g^4 \left((4 \kappa+2) M_V^2+3 \kappa M_W^2\right)+\lambda_{HV} ^2 M_W^2\right)}{32 \pi  g^2 M_V^4}-\frac{\left(3 \alpha_1+4 \alpha_2+3 \alpha_3+4 \kappa^2 g^2\right)s^2 }{96 \pi  M_V^4}.
\end{equation*}

\underline{$V^{++}V^{-}\to V^{++} V^{-}$}
\begin{equation*}
    a_0(s)=\frac{s \left(\left(3 \alpha_1+2 \alpha_2+3 \alpha_3\right) g^2 M_V^2+2 \kappa g^4 \left((4 \kappa+2) M_V^2+3 \kappa M_W^2\right)-\lambda_{HV} ^2 M_W^2\right)}{32 \pi  g^2 M_V^4}-\frac{ \left(2 \alpha_1+\alpha_2+2 \alpha_3+8 \kappa^2 g^2\right)s^2 }{96 \pi  M_V^4}.
\end{equation*}

\underline{$V^{++}V^{0}\to V^{++} V^{0}$}
\begin{equation*}
    a_0(s)=\frac{s \left(\left(3 \alpha _1+2 \alpha _2+3 \alpha _3\right) g^2 M_V^2-\lambda_{HV} ^2 M_W^2\right)}{32 \pi  g^2 M_V^4}-\frac{\left(2 \alpha _1+\alpha _2+2 \alpha _3\right) s^2}{96 \pi  M_V^4}.
\end{equation*}

\underline{$V^{+}V^{0}\to V^{+} V^{0}$}
\begin{equation*}
    a_0(s)=\frac{s \left(3 g^4 \kappa  \left((4 \kappa +2) M_V^2+3 \kappa  M_W^2\right)+\left(3 \alpha _1+2 \alpha _2+3 \alpha _3\right) g^2 M_V^2-\lambda_{HV} ^2 M_W^2\right)}{32 \pi  g^2 M_V^4}-\frac{\left(2 \alpha _1+\alpha _2+2\alpha _3+12 g^2 \kappa ^2\right)s^2}{96 \pi  M_V^4}.
\end{equation*}

\underline{$V^{++}V^{+}\to V^{++} V^{+}$}
\begin{equation*}
    a_0(s)=\frac{s \left(\left(3 \alpha _1+2 \alpha _2+3 \alpha _3\right) g^2 M_V^2-\lambda_{HV} ^2 M_W^2\right)}{32 \pi  g^2 M_V^4}-\frac{\left(2 \alpha _1+\alpha _2+2 \alpha _3\right) s^2}{96 \pi  M_V^4}.
\end{equation*}

\underline{$V^{++}V^{++}\to V^{++} V^{++}$}
\begin{equation*}
    a_0(s)=\frac{s \left(3 \alpha_1+2 \alpha_2+3 \alpha_3-\frac{4 \kappa g^4 \left((4 \kappa+2) M_V^2+3 \kappa M_W^2\right)+\lambda_{HV} ^2 M_W^2}{g^2 M_V^2}\right)}{16 \pi  M_V^2}-\frac{\left(2 \alpha_1+\alpha_2+2 \alpha_3-16 \kappa^2 g^2\right)s^2 }{48 \pi  M_V^4}.
\end{equation*}

\underline{$V^{+}V^{+}\to V^{+} V^{+}$}
\begin{equation*}
    a_0(s)=\frac{s \left(3 \alpha_1+2 \alpha_2+3 \alpha_3-\frac{\kappa g^4 \left((4 \kappa+2) M_V^2+3 \kappa M_W^2\right)+\lambda_{HV} ^2 M_W^2}{g^2 M_V^2}\right)}{16 \pi  M_V^2}-\frac{\left(2 \alpha_1+\alpha_2+2 \alpha_3-4 \kappa^2 g^2\right)s^2 }{48 \pi  M_V^4}.
\end{equation*}

\underline{$V^0V^0\to V^0V^0$}
\begin{equation*}
    a_0(s)=\frac{s\left(\alpha_1+\alpha_2+\alpha_3\right) }{4 \pi  M_V^2}-\frac{5 \left(\alpha_1+\alpha_2+\alpha_3\right) s^2}{96 \pi  M_V^4}.
\end{equation*}

\underline{$V^{++}V^{-}\to V^{+}V^0$}
\begin{equation*}
    a_0(s)=\frac{\sqrt{\frac{3}{2}} g^2 \kappa   \left((4 \kappa +2) M_V^2+3 \kappa  M_W^2\right)s}{16 \pi  M_V^4}+\frac{g^2 \kappa ^2 s^2}{4 \sqrt{6} \pi  M_V^4}.
\end{equation*}

\underline{$ZZ\to V^0 V^0$}
\begin{equation*}
    a_0(s)=\frac{\lambda_{HV}  s}{32 \pi  M_V^2}.
\end{equation*}

\underline{$Z V^0\to Z V^0$}
\begin{equation*}
    a_0(s)=-\frac{\lambda_{HV}  s}{64 \pi  M_V^2}.
\end{equation*}
\\

\normalsize
The inclusion of $\xi$ as a `gauge-fixing-like' parameter for the field $V$ ensures a good behavior at high energies for channels involving the exchange of a Higgs boson or a $V$-particle. On the other hand, the contact diagram $hhVV$ still exhibits a energy dependence on $s$ at this scale.
\\

\footnotesize
\underline{$hV\to hV$ / $hh\to VV$}
\begin{equation*}
    a_0(s)=-\frac{\lambda_{HV} s}{64 \pi  M_V^2}.
\end{equation*}

\underline{$ZZ\to V^+ V^-$}
\begin{equation*}
    a_0(s)=\frac{s \left(3 g^2 \cos^2{(\theta_W)} \left(\cos^2{(\theta_W)} M_V^2 \left(\frac{4}{\xi }+3\right)+4 M_W^2 \left(\kappa +\frac{1}{\xi }\right)\right)+2 \lambda_{HV}  M_W^2\right)}{64 \pi  M_V^2 M_W^2}-\frac{g^2 \cos^4{(\theta_W)} s^2}{16 \pi \xi M_V^2   M_W^2}.
\end{equation*}

\underline{$ZZ\to V^{++} V^{--}$}
\begin{equation*}
    a_0(s)=\frac{s \left(6 g^2 \cos^2{(\theta_W)} \left(\cos^2{(\theta_W)} M_V^2 \left(\frac{4}{\xi }+3\right)+4 M_W^2 \left(\kappa +\frac{1}{\xi }\right)\right)+\lambda_{HV}  M_W^2\right)}{32 \pi  M_V^2 M_W^2}-\frac{g^2 \cos^4{(\theta_W)} s^2}{4 \pi \xi M_V^2   M_W^2}.
\end{equation*}

\underline{$Z V^+\to Z V^+$}
\begin{equation*}
    a_0(s)=\frac{s \left(g^2 \cos^2{(\theta_W)} \left(\cos^2{(\theta_W)} M_V^2 \left(-\frac{8}{\xi ^2}+\frac{44}{\xi }-9\right)+4 W^2 \left(\frac{\frac{2}{\xi }+7}{\xi }-3 \kappa \right)\right)-2 \lambda_{HV}  M_W^2\right)}{128 \pi  M_V^2 M_W^2}-\frac{g^2 \cos^4{(\theta_W)} s^2}{8 \pi \xi M_V^2   M_W^2}.
\end{equation*}

\underline{$Z V^{++}\to Z V^{++}$}
\begin{equation*}
    a_0(s)=-\frac{s \left(2 g^2 \cos^2{(\theta_W)} \left(\cos^2{(\theta_W)} M_V^2 \left(\frac{8}{\xi ^2}-\frac{44}{\xi }+9\right)+4 W^2 \left(3 \kappa -\frac{\frac{2}{\xi }+7}{\xi }\right)\right)+\lambda_{HV}  M_W^2\right)}{64 \pi  M_V^2 M_W^2}-\frac{g^2 \cos^4{(\theta_W)} s^2}{2 \pi \xi M_V^2  M_W^2}.
\end{equation*}

\underline{$W^+V^{++}\to W^+V^{++}$}
\begin{equation*}
    a_0(s)=-\frac{g s \left(2 g M_V^2 \left(\frac{4}{\xi }+1\right)+4 g M_W^2 \left(\kappa +\frac{2}{\xi }-2\right)+\lambda_{HV}  v M_W\right)}{128 \pi  M_V^2 M_W^2}+\frac{g^2 \left(\frac{1}{\xi }-2\right) s^2}{48 \pi  M_V^2 M_W^2}.
\end{equation*}

\underline{$W^+V^{+}\to W^+V^{+}$}
\begin{equation*}
    a_0(s)=\frac{g s \left(g M_V^2 \left(25-\frac{28}{\xi }\right)+2 g M_W^2 \left(11 \kappa -\frac{14}{\xi }+6\right)-\lambda_{HV}  v M_W\right)}{128 \pi  M_V^2 M_W^2}+\frac{g^2 \left(\frac{3}{\xi }-2\right) s^2}{32 \pi  M_V^2 M_W^2}.
\end{equation*}

\underline{$V^0W^+ \to V^0W^+$}
\begin{equation*}
    a_0(s)=\frac{g s \left(3 g \left(M_V^2 \left(\frac{4 \left(\frac{2}{\xi }-5\right)}{\xi }+13\right)+4 M_W^2 \left(3 \kappa -\frac{2}{\xi ^2}-\frac{1}{\xi }+1\right)\right)-\lambda_{HV}  v M_W\right)}{128 \pi  M_V^2 M_W^2}+\frac{g^2 \left(\frac{2}{\xi }-1\right) s^2}{16 \pi  M_V^2 M_W^2}.
\end{equation*}

\underline{$V^+W^{-}\to V^+W^{-}$}
\begin{equation*}
    a_0(s)=\frac{g s \left(8 g M_V^2 \left(\frac{8}{\xi }-7\right)+2 g M_W^2 \left(-29 \kappa +\frac{32}{\xi }+4\right)-\lambda_{HV}  v M_W\right)}{128 \pi  M_V^2 M_W^2}-\frac{g^2 \left(\frac{25}{\xi }+4\right) s^2}{96 \pi  M_V^2 M_W^2}.
\end{equation*}

\underline{$W^{+}V^{--}\to W^{+}V ^{--}$}
\begin{equation*}
    a_0(s)=-\frac{g s \left(4 g \left(M_V^2 \left(\frac{4}{\xi }-7\right)+M_W^2 \left(\frac{4}{\xi }-7 \kappa \right)\right)+\lambda_{HV}  v M_W\right)}{128 \pi  M_V^2 M_W^2}+\frac{g^2 s^2}{16 \pi \xi M_V^2  M_W^2}.
\end{equation*}

\underline{$W^{-}V^{++}\to W^+V^0$}
\begin{equation*}
    a_0(s)=\frac{\sqrt{\frac{3}{2}} g^2 s \left(M_V^2 \left(\frac{12}{\xi }-13\right)-4 M_W^2 \left(3 \kappa -\frac{3}{\xi }+1\right)\right)}{64 \pi  M_V^2 M_W^2}+\frac{g^2 \left(\frac{2}{\xi }-1\right) s^2}{8 \sqrt{6} \pi  M_V^2 M_W^2}.
\end{equation*}

\underline{$W^+W^- \to V^0V^0$}
\begin{equation*}
    a_0(s)=\frac{g s \left(\lambda_{HV}  v M_W-3 g \left(M_V^2 \left(\frac{4}{\xi }+13\right)+4 M_W^2 \left(3 \kappa +\frac{1}{\xi }+1\right)\right)\right)}{64 \pi  M_V^2 M_W^2}+\frac{g^2 \left(\frac{1}{\xi }+2\right) s^2}{16 \pi  M_V^2 M_W^2}.
\end{equation*}

\underline{$W^+W^- \to V^+V^-$}
\begin{equation*}
    a_0(s)=\frac{g s \left(2 \lambda_{HV}  v M_W-5 g \left(M_V^2 \left(\frac{4}{\xi }+13\right)+4 M_W^2 \left(3 \kappa +\frac{1}{\xi }+1\right)\right)\right)}{128 \pi  M_V^2 M_W^2}+\frac{5 g^2 \left(\frac{1}{\xi }+2\right) s^2}{96 \pi  M_V^2 M_W^2}.
\end{equation*}

\underline{$W^+W^- \to V^{++}V^{--}$}
\begin{equation*}
    a_0(s)=\frac{g s \left( \lambda_{HV} v M_W - g \left( M_V^2 \left(\frac{4}{\xi }+19\right)+2 M_W^2 \left(6 \kappa +\frac{2}{\xi }+5\right)\right)\right)}{64 \pi  M_V^2 M_W^2}+\frac{g^2 \left(\frac{4}{\xi }+17\right) s^2}{192 \pi  M_V^2 M_W^2}.
\end{equation*}

\normalsize
\section{Potentials for Sommerfeld Enhancement }
\label{sec: Appendix Sommerfeld}
\subsection{Cubic terms}
In Sec.~\ref{subsec: Indirect}, equation~\eqref{Schrodinger}, we introduce the Schrödinger equation for non-relativistic DM annihilation.
The calculation of the potential $V(r)$, related to trilinear interactions in the Lagrangian~\eqref{final_lagrangian}, giving rise to Coulomb and Yukawa potentials, require computing the following effective action:
\begin{align}
    \label{effeaction1}
    S_{eff,1}\sim\int d^{4}x~ \left[J_{\mu}^{W^{+}}(x)W^{+\mu}(x)+J_{\mu}^{W^{-}}(x)W^{-\mu}(x)+J_{\mu}^{Z}(x)Z^{\mu}(x)+J_{\mu}^{A}(x)A^{\mu}(x)\right].
\end{align}

After integrating out the gauge bosons in the path integral, we obtain the effective action dependent on the currents. 
The conserved currents are calculated from the cubic terms of the Lagrangian \eqref{final_lagrangian} \cite{Chowdhury:2016mtl}. The previous expression in terms of the currents is
\begin{align}\label{acceffnon1}
S_{eff,1}=-\int\dfrac{d^{4}xd^{3}y}{8\pi|\vec{x}-\vec{y}|}&(J_{A}^{0}(x)J_{A}^{0}(x^{0},\vec{y})+J_{Z}^{0}(x)J_{Z}^{0}(x^{0},\vec{y})e^{-M_{Z}|\vec{x} -\vec{y}|}\nonumber\\ &+2J_{W^{+}}^{0}(x)J_{W^{-}}^{0}(x^{0},\vec{y})e^{-M_{W}|\vec{x}-\vec{y}|}).
\end{align}
The first term in equation \eqref{acceffnon1} accounts for the Coulomb interaction, while the remaining terms describe Yukawa interactions associated with the exchange of $W$ and $Z$ bosons.

We also need to perform the non-relativistic limit of the exotic fields \cite{Abe:2021mry}:
\begin{equation*}
    \label{vo}
    V_{i}^{0}(x)=\dfrac{1}{\sqrt{2M_{V}}}[v_{0}(\vec{x})\epsilon_{i}(x)e^{iM_{V}t}+v_{0}^{\dagger}(\vec{x})\epsilon_{i}^{\ast}(x)e^{-iM_{V}t}],
\end{equation*}
\begin{equation*}
    \label{v+}
    V_{i}^{+}(x)=\dfrac{1}{\sqrt{2M_{V}}}[v_{+}(\vec{x})\epsilon_{i}(x)e^{iM_{V}t}+v_{-}^{\dagger}(\vec{x})\epsilon_{i}^{\ast}(x)e^{-iM_{V}t}],
\end{equation*}
\begin{equation*}
    \label{v-}
    V_{i}^{-}(x)=\dfrac{1}{\sqrt{2M_{V}}}[v_{+}^{\dagger}(\vec{x})\epsilon_{i}^{\ast}(x)e^{-iM_{V}t}+v_{-}(\vec{x})\epsilon_{i}(x)e^{iM_{V}t}],
\end{equation*}
\begin{equation*}
    \label{v++}
    V_{i}^{++}(x)=\dfrac{1}{\sqrt{2M_{V}}}[v_{++}(\vec{x})\epsilon_{i}(x)e^{iM_{V}t}+v_{--}^{\dagger}(\vec{x})\epsilon_{i}^{\ast}(x)e^{-iM_{V}t}],
\end{equation*}
\begin{equation*}
    \label{v--}
    V_{i}^{--}(x)=\dfrac{1}{\sqrt{2M_{V}}}[v_{++}^{\dagger}(\vec{x})\epsilon_{i}^{\ast}(x)e^{-iM_{V}t}+v_{--}(\vec{x})\epsilon_{i}(x)e^{iM_{V}t}].
\end{equation*}
The potential is then expressed as
\begin{equation*}
    \label{acNR}
    S_{eff,1}^{NR}=-\int d^{4}xd^{3}y~s(x,\vec{y})^{\dagger}V_{gauge}(|\vec{x}-\vec{y}|)s(x,\vec{y}),
\end{equation*} 
where $s(x,\vec{y})$ is given by
\begin{equation}
    \label{vectornonrel}
    s(x,\Vec{y})=
    \begin{pmatrix}
        v_{+}(x)v_{-}(x^{0},\Vec{y})\\
        \dfrac{v_{0}(x)v_{0}(x^{0},\Vec{y})}{\sqrt{2}}\\
        v_{++}(x)v_{--}(x^{0},\Vec{y})
    \end{pmatrix},   
\end{equation}
We need to consider only the vector \eqref{vectornonrel}, constructed with neutral pairs of fields, to calculate DM annihilation. The non-relativistic potential of the gauge fields is then given by
\begin{align}
    \label{Vgauge1}
    V_{\text{gauge}}(r)=-\dfrac{g^{2}}{32\pi\xi^{2} r}
    \begin{pmatrix}
        2(c_{W}^{2}+s_{W}^{2}e^{-M_{Z}r}) & 3\sqrt{2}e^{-M_{W}r} & 2e^{-M_{W}r} \\
        3\sqrt{2}e^{-M_{W}r} & 0 & 0\\
        2e^{-M_{W}r} & 0 & 8(c_{W}^{2}+s_{W}^{2}e^{-M_{Z}r})
    \end{pmatrix},
\end{align}
where $r=|\vec{x}-\vec{y}|$.
Now, from the interactions of the quintuplet with the Higgs doublet,
\begin{align}
    \label{lag cub H}
    \mathcal{L}_{hV}=\frac{\lambda_{HV}^{2}}{2}(\Phi^\dagger \Phi)(V_{\mu}^\dagger V^{\mu}), 
\end{align}
we can write the corresponding non-relativistic potential as
\begin{align}
    \label{Vhiggs1}
    V_{\text{Higgs}}(r)=\dfrac{v^{2}\lambda_{HV}^{2}(e^{-M_{h}r}+2e^{-M_{Z}r})}{8\pi M_{V}^{2}r}
    \begin{pmatrix}
        1 & 1/2 & 1\\
        1/2 & 4 & 1\\
        1 & 1 & 1
    \end{pmatrix}.
\end{align}

\subsection{Cuartic terms}
Low-velocity vector DM particles can annihilate into SM particles. 
In non-relativistic quantum mechanics, these annihilations are represented by an imaginary potential, or absorptive term, in the Schrödinger equation, accounting for the creation or destruction of particles.

The effective action for absorptive terms is
\begin{align}\label{seff2}
S_{eff,2}&=-i Tr\log\left(\Delta^{-1}\mathcal{U}\Delta^{-1}\mathcal{U}\right)\nonumber\\
&=\frac{i}{2\pi}\int d^{4}x~ Tr\left[\begin{pmatrix}
U_{AA}(x)&U_{AZ}(x)&U_{AW}(x)&U_{AW}^{\ast}(x)\\
U_{AZ}(x)&U_{ZZ}(x)&U_{ZW}(x)&U_{ZW}^{\ast}(x)\\
U_{AW}^{\ast}(x)&U_{ZW}^{\ast}(x)&U_{WW}(x)&0\\
U_{AW}^{\ast}(x)&U_{ZW}^{\ast}(x)&0&U_{WW}^{\ast}(x)\\
\end{pmatrix}\right]^{2},
\end{align}
This expression is easy to understand in a diagrammatic way: the matrix $\Delta$ represents the propagators of the gauge fields, and each $\mathcal{U}$ represents a vertex involving two gauge fields and two vector DM fields \cite{Chowdhury:2016mtl,GarciaCely:2014jha}.

In terms of the vector $\vec{s}$, this effective action defines absorptive terms by
\begin{equation}
    S_{eff,2}=2i\int d^{4}x d^{3}y~s(x,\vec{y})^{\dagger}\Gamma_\text{{gauge}}\delta^{3}(|x-y|)s(x,\vec{y})  
\end{equation}
We can express this effective action in terms of the new fields as:
\begin{equation}
    \label{sQs1}
    S_{eff,2}=\dfrac{i}{2\pi}\int~d^{4}x\left(U_{AA}(x)^{2}+U_{ZZ}(x)^{2}+2U_{WW}(x)^{2}+2U_{AZ}(x)^{2} \right)
\end{equation}
where the $U$ terms  are written in terms of the exotic vector fields as
\begin{equation*}
    U_{ZZ}(x)=\dfrac{g^{2}}{4\xi}s_{W}^{2}(V^{+\mu}V^{-\nu}+4V^{++\mu}V^{--\nu}), 
\end{equation*}
\begin{equation*}
    U_{AA}(x)=\dfrac{g^{2}}{4\xi}c_{W}^{2}(V^{+\mu}V^{-\nu}+4V^{++\mu}V^{--\nu}),
\end{equation*}
\begin{equation*}
    U_{WW}(x)=\dfrac{g^{2}}{8\xi}(5V^{+\mu}V^{-\nu}+3V^{0\mu}V^{0\nu}+2V^{++\mu}V^{--\nu}),
\end{equation*}
\begin{equation*}
    U_{AZ}(x)=\dfrac{g^{2}}{4\xi}c_{W} s_{W}(V^{+\mu}V^{-\nu}+4V^{++\mu}V^{--\nu}),
\end{equation*}
The various contributions to the absorptive term are
\begin{align}
    \label{gWWJ=0}
    \Gamma_{WW}^{J=0}=\dfrac{g^{4}}{32\sqrt{3}\pi\xi^{2} m^{2}}
    \begin{pmatrix}
        25&15&10\\
        15&0&6\\
        6&10&4
    \end{pmatrix},
\end{align}
\begin{align}
    \label{gWWJ=2}
    \Gamma_{WW}^{J=2}=\dfrac{g^{4}}{64\pi\xi^{2} m^{2}}\left(\dfrac{1+\cos(2\theta)}{2\sqrt{6}}+\sin^{2}\theta\right)
    \begin{pmatrix}
        25&15&10\\
        15&0&6\\
        6&10&4
    \end{pmatrix},
\end{align}
\begin{align}
    \label{gZZJ=0}
    \Gamma_{ZZ}^{J=0}=\dfrac{g^{4}s_{W}^{4}}{32\sqrt{3}\pi \xi^{2}m^{2}}
    \begin{pmatrix}
        1&0&4\\
        0&0&0\\
        4&0&16
    \end{pmatrix},
\end{align}
\begin{align}
    \label{gZZJ=2}
    \Gamma_{ZZ}^{J=2}=\dfrac{g^{4}s_{W}^{4}}{64\pi\xi^{2} m^{2}}\left(\dfrac{1+\cos(2\theta)}{2\sqrt{6}}+\sin^{2}\theta\right)
    \begin{pmatrix}
        1&0&4\\
        0&0&0\\
        4&0&16
    \end{pmatrix},
\end{align}
\begin{align}
    \label{gAAJ=0}
    \Gamma_{AA}^{J=0}=\dfrac{g^{4}c_{W}^{4}}{32\sqrt{3}\pi\xi^{2} m^{2}}\begin{pmatrix}
    1&0&4\\
    0&0&0\\
    4&0&16
    \end{pmatrix},
\end{align}
\begin{align}
    \label{gAAJ=2}
    \Gamma_{AA}^{J=2}=\dfrac{g^{4}c_{W}^{4}}{64\pi\xi^{2} m^{2}}\left(\dfrac{1+\cos(2\theta)}{2\sqrt{6}}+\sin^{2}\theta\right)
    \begin{pmatrix}
        1&0&4\\
        0&0&0\\
        4&0&16
    \end{pmatrix},
\end{align}
\begin{align}
    \label{gAZJ=0}
    \Gamma_{AZ}^{J=0}=\dfrac{g^{4}c_{W}^{2}s_{W}^{2}}{8\sqrt{3}\pi\xi^{2} m^{2}}
    \begin{pmatrix}
        1&0&4\\
        0&0&0\\
        4&0&16
    \end{pmatrix},
\end{align}
\begin{align}
    \label{gAZJ=2}
    \Gamma_{AZ}^{J=2}=\dfrac{g^{4}c_{W}^{2}s_{W}^{2}}{16\pi\xi^{2} m^{2}}\left(\dfrac{1+\cos(2\theta)}{2\sqrt{6}}+\sin^{2}\theta\right)\begin{pmatrix}
    1&0&4\\
    0&0&0\\
    4&0&16
\end{pmatrix}.
\end{align}

Analogously, we now consider the quartic terms in the Higgs portal interaction term of \eqref{lag cub H}, which gives rise to the following contribution to the absorptive term:
\begin{align}
    \label{Ghiggs}
    \Gamma_{Higgs}=\dfrac{25\lambda_{HV}^{2}}{16\pi M_{V}^{2}}
    \begin{pmatrix}
        1 & 1/2 & 1\\
        1/2 & 4 & 1\\
        1 & 1 & 1
    \end{pmatrix}.
\end{align}

\end{document}